\PassOptionsToPackage{frozencache,cachedir=minted-cache}{minted}
\PassOptionsToPackage{usenames,dvipsnames}{xcolor}
\documentclass[acmsmall,screen,nonacm]{acmart}
\AtBeginDocument{}

\usepackage{rcompcerto}
\usepackage{lstcoq}

\begin{document}

\title{Verified Implementation of the Next-gen Borrow Checking Algorithm for Rust via Abstract Interpretation}

\author{Jinhua Wu}
\orcid{0000-0001-5812-053X}             \affiliation{
\department{John Hopcroft Center for Computer Science, School of Electronic Information and Electrical Engineering}              \institution{Shanghai Jiao Tong University}            \country{China}                    }
\email{jinhua.wu@sjtu.edu.cn}          

\author{Yuting Wang}
\authornote{Corresponding author}          \orcid{0000-0003-3990-2418}             \affiliation{
\department{John Hopcroft Center for Computer Science, School of Electronic Information and Electrical Engineering}              \institution{Shanghai Jiao Tong University}            \country{China}                    }
\email{yuting.wang@sjtu.edu.cn}          

\author{Zhemin Chen}
\orcid{0009-0006-2582-1878}             \affiliation{
\department{Software Institute}              \institution{Nanjing University}            \country{China}                    }
\email{zheminchen@smail.nju.edu.cn>}

\begin{abstract}

Borrow checking is a central mechanism for Rust to ensure temporal memory safety. Its correctness is based on the intuition that the Aliasing XOR Mutability discipline guarantees that programs without borrow errors inherently could not access invalid references (i.e., pointers). 
However, existing work on formalizing borrow checking treats it either as a component of type systems or as symbolic execution describing resource transfer. 
None provides a direct explanation of the correctness of borrow checking \emph{algorithms} such as Non-Lexical Lifetimes (NLL) or Polonius in the official Rust compiler that function more like static pointer analysis.

We take a first step toward solving this problem by formalizing
Polonius, the next-generation borrow-checking algorithm for the Rust
compiler, on a core subset of Rust. We prove its correctness by
showing that any program accepted by Polonius does not access invalid
references at runtime.
The key observation is that, when borrow checking is performed \emph{lazily}, Polonius could be treated as a forward abstract interpretation on abstract domains recording pointer aliasing (borrowing) relations.
By providing a concrete semantics that formally describes the Aliasing XOR Mutability discipline, we prove the soundness of this abstract interpretation, from which we derive memory safety of checked programs.
This formulation of Polonius is implemented and formally verified in the abstract interpretation framework of CompCert. The correctness proofs are generated by an agentic AI under human guidance using paper proofs as proof outline.
To the best of our knowledge, this is the first comprehensive formalization and verification of a borrow checking algorithm as manifested in the Rust compiler.
  
\end{abstract}

\maketitle

\section{Introduction}\label{sec:intro}

Memory safety problems caused by accessing invalid pointers or references are the most common vulnerabilities in software~\cite{microsoft-memory-safety,chromium-memory-safety}. To mitigate those problems, memory safe languages are increasingly favored over unsafe ones (e.g., C/C++). Among them the most prominent is Rust~\cite{rust}.
Rust employs a strong type system and ownership semantics to ensure both \emph{temporal} (e.g., no dereference of dangling pointers) and \emph{spatial} memory safety (e.g., no out-of-bounds access).
For the former, the central idea is to enforce the so-called \emph{Aliasing XOR Mutability} discipline (i.e., there can be either only one \emph{mutable} reference or multiple \emph{shared} references to a value at any execution point) to rule out potentially unsafe uses of references. We shall abbreviate this discipline as \emph{AXM}.

Borrow checking is the static analysis for Rust to ensure memory safety by enforcing AXM.
Consider the following Rust code snippet which manipulates heap memory owned by \code{Box} structures:
\begin{minted}[xleftmargin=1.5em,mathescape=true,escapeinside=||]{rust}
let mut v = Box::new(1);
let mut r = &mut *v;
v = Box::new(2);    // error: cannot assign to v because it is borrowed
println!("{}", *r); // borrow later used here
\end{minted}
If the code is compiled and executed without any safety checking, a use-after-free error occurs at line 4 because the heap memory originally bound to \mcode{v} has been deallocated after line 3, making \mcode{r} a dangling reference.
The Rust compiler rejects this program as its borrow checker detects a violation of AXM at line 3: \code{*v} is mutably borrowed by \code{r} until line 4 and therefore its prefixes (i.e., \mcode{v} and \mcode{*v}) could not be mutated before line 4.
In summary, borrow checking \emph{statically} examines the possible violations of AXM, thereby preventing any potential access to invalid references at \emph{runtime}.

\begin{figure}[t]
\begin{minted}[xleftmargin=1.5em,mathescape=true,escapeinside=||]{rust}
let mut v = Box::new(1);          //     NLL    |      Polonius      
let mut x: &'x mut i32 = &mut *v; //   *v live  |      'x:{*v}       
let mut y: &'y mut i32 = &mut *x; // *v,*x live | 'x:{*v},'y:{*v,*x} 
*x = 2; // Error: cannot access *x because it is borrowed
*y = 3; // borrow later used here
\end{minted}
\caption{An example to illustrate NLL and Polonius}
\label{fig:intro-exm}
\end{figure}

 In the official compiler of Rust, i.e., \mcode{rustc}, borrow checking is implemented as an algorithm that solves memory safety constraints and checks if the solution violates AXM~\cite{rust-nll,polonius-alias}. To deal with complicated programming language features in Rust such as storing references in data structures, reborrowing of existing references, passing and returning references for function calls, etc., borrow checking algorithms are highly sophisticated and continuously evolving. Two versions of borrow checkers are currently implemented in \mcode{rustc}: \emph{Non-Lexical Lifetimes} (NLL)~\cite{rust-nll} and \emph{Polonius}~\cite{polonius-alias, Polonius-update}. The former has been the default borrow checker for a long time and the latter is an experimental one that employs a more accurate borrow analysis and will replace NLL in the near future.

In both NLL and Polonius, the key to ensuring that only valid references are accessed lies in analyzing which \emph{places} (Rust's terminology for memory locations holding values, similar to \emph{l-values} in C/C++) may be borrowed at each program point, based on which violation of AXM is then detected. 
This is made possible by annotating types of references with generic variables called \emph{regions}.
Consider the example in~\figref{fig:intro-exm} where \ticktt{x} and \ticktt{y} are regions in the types of references \mcode{x} and \mcode{y}, respectively. 
In NLL, a region is interpreted as \emph{lifetime}, i.e., a span of code
over which a reference and its borrowed places remain alive (i.e., they will be used later).
For instance, the lifetime denoted by \ticktt{x} spans over lines 2-4. At line 3, a \emph{reborrow} of \mcode{*v} by \mcode{y} happens through dereferencing \mcode{x}. The lifetime of \ticktt{y} spans over lines 3-5, making its borrowed place \code{*x} alive up to line 5. Because \mcode{*x} is still mutably borrowed at line 4, it cannot be assigned, resulting in a borrow checking error.
By contrast, Polonius interprets a region $r$ as an abstract pointer variable that stores a set $s$ of borrowed places, which we shall write as $r:s$. At any point of execution, this set (over-)approximates all the possible places the reference variable could borrow. For example, at line 2,
$\mcode{\ticktt{x}:\rawset{*v}}$ indicates that \mcode{x} may
borrow \mcode{\deref v} as \ticktt{x} is the abstract pointer
representing \mcode{x}. At line 3, the places borrowed by \mcode{y} are additionally recorded in $\mcode{\ticktt{y}:\rawset{*v,*x}}$.
Using this borrowing information, Polonius checks if, at any point of execution, the instruction being executed violates AXM w.r.t. borrowed places contained in \emph{live} regions.
For example, because \mcode{y} (hence its region \ticktt{y}) is alive up to line 5 and it borrows \mcode{*x}, assignment to \mcode{*x} at line 4 violates AXM. 
As we can see, Polonius captures the borrow relations for \emph{individual references} at all points of execution, and is therefore more accurate than NLL which  only records live places \emph{as a whole}. This enables Polonius to accept a wider range of safe programs, a main reason that it is chosen as the next-gen borrow checking algorithm for \code{rustc}.~\footnote{\figref{fig:intro-exm} also illustrates the conservative nature of borrow checkers. Although line 4 does not actually cause a memory error, the program is still rejected by \mcode{rustc}.  Inaccuracy is the price we pay for the ability to statically check memory safety.}

Considering the central role borrow checking algorithms play in the Rust compiler, it is highly desirable to formally verify their correctness.
However, despite the impressive efforts the PL community puts into researching Rust, there is not yet successful attempt along this line.
Instead, existing work formalizes borrow checking either as an ownership type system for Rust (e.g., RustBelt~\cite{RustBelt}, Oxide~\cite{weiss2019oxide}, Pearce~\cite{light-rust}), or as symbolic execution with resource transfers (e.g., Aeneas~\cite{Aeneas}), which deviates from the region-based static analysis adopted by NLL or Polonius.
Verifying the correctness of borrow checking algorithms in the Rust compiler is still an open~problem.

In this paper, we investigate a solution to this open problem by developing a verified implementation of Polonius.
Before delving into the solution, we first explain why the problem is difficult.

\myparagraph{Challenge 1: Formalization of Borrow Checking Algorithms Amenable to Verification}
To formally prove borrow checkers correct, we must first give them a formal description in theorem provers. However, this is difficult because of the ad-hoc and complex implementation of borrow checkers. In \mcode{rustc}, NLL and Polonius employ sophisticated constraint generation and solving techniques~\cite{rust-subtyping,rust-nll,polonius-alias} which are entangled with other kinds of safety checking such as type checking, liveness checking, and module-level analysis. Although there is documentation like RFCs, blog posts, and even a rule-based formulation in Datalog (at least partially for Polonius~\cite{rust-polonius}), it is not clear at all if there exists a formulation of NLL or Polonius amenable to verification.

\myparagraph{Challenge 2: Semantics Suitable for Verifying Borrow Checking Algorithms}
Borrow checkers work on MIR, an intermediate language represented as control-flow graphs (CFG). 
For verification, we need to give this IR a formal semantics that faithfully describes when memory errors occur at runtime, so that we can prove their absence under successful borrow checking.
More importantly, we need to connect this low-level semantics with a more higher-level semantics that captures the essence of borrow checkers. For Polonius, this essence is its \emph{alias analysis} nature. For example, the relation between borrowed places $\mcode{\ticktt{y}:\rawset{*v,*x}}$ in~\figref{fig:intro-exm} actually implies that \mcode{*v}, \mcode{*x} and \mcode{*y} are aliases of each other. It is exactly by analyzing this alias information that Polonius is able to enforce AXM, hence to guarantee memory safety.

\myparagraph{Challenge 3: Proving the Correctness of Borrow Checking Algorithms}
Intuitively, borrow checkers in \mcode{rustc} are correct, i.e., if they could not detect any borrow error, then the program does not contain any uses of invalid references.
However, to formally prove this intuition is a significant challenge. First and foremost, the actual form of AXM is more complicated because of \emph{pointer assignments}, which is another term for \emph{reborrowing}. For example, the reborrowing at line 3 in~\figref{fig:intro-exm} is actually desugared from the following pointer assignment by \mcode{rustc}~\cite{rust-auto-reborrow}:
\begin{minted}[xleftmargin=1.5em,mathescape=true,escapeinside=||]{rust}
 let mut y: &'y mut i32 = x;
\end{minted}
In real-world software, pointer assignments (hence reborrowing) are  ubiquitous. They create multiple mutable references to values, which easily breaks AXM. Therefore, Rust actually allows multiple mutable references to coexist provided that their uses are \emph{well-nested} like a stack. For example, if we switch lines 4 and 5 in~\figref{fig:intro-exm}, then the modified program passes borrow checking because \mcode{x} and \mcode{y} are used in a first-defined-last-used fashion. 
As a result, the more complex \emph{stacked AXM discipline}, i.e., that \textbf{only well-nested memory accesses according to the stack ordering are allowed}, must be taken into account when proving correctness.

Second, pointer manipulations in real-world code further complicate the proof. For example, the assignment between nested references (i.e., pointers to pointers) causes aliases of locations to merge; reference arguments in functions and struct fields could be annotated with generic regions called \emph{lifetime parameters} which extend the lifetime of references indefinitely.
It is unclear how accesses to references could be proven safe in presence of these complications.

Finally, our ultimate goal is to derive memory safety from borrow checking. Although there is the folklore that AXM implies absence of accessing invalid references, it is not clear what exactly is a formal definition of this implication relation.
Even if we know this formal definition, proving it holds for stacked AXM under complicated pointer manipulation is still a huge undertaking.

\subsection{Our Contributions}
\label{ssec:contrib}
In this paper, we develop a verified implementation of the Polonius borrow checking algorithm for a core subset of Rust. The key observation is that, Polonius coincides with an abstract interpretation of \emph{dataflow of pointer aliases} when detection of borrow errors is done \emph{lazily}. We adopt the traditional fixed-point solver (i.e., Kildall's algorithm) to implement this abstract interpretation as a formal algorithm in Rocq and prove its soundness with respect to a concrete semantics we call \emph{borrow-aware semantics}. By further proving that, when there is no  borrow checking error, borrow-aware semantics enforces stacked AXM discipline, we derive memory safety for checked programs.

To the best of our knowledge, this is the first time a borrow checking algorithm in the Rust compiler is formally implemented and verified. We summarize our contributions as follows.
\begin{itemize}[leftmargin=*]
	
\item \textbf{Polonius as Abstract Interpretation.}
We begin with an understanding of the official Polonius as derivation rules that capture its constraint generation and solving mechanisms. We observe that, when borrow errors are detected lazily instead of eagerly, these rules function like a forward propagation of aliasing information for pointers. Based on this observation, for a core subset of MIR, we develop an abstract semantics for interpreting Polonius as a variant of \emph{pointer analysis}~\cite{pointer-survey,andersen1994program}.  We propose a novel abstract domain to track aliasing information in presence of complex pointer manipulation like reborrowing.
This formulation of Polonius as abstract interpretation is implemented as a formal algorithm and is naturally amenable to verification.

\item \textbf{Borrow-Aware Semantics Capturing AXM and Memory Safety.}
We propose \emph{borrow-aware semantics} for our Rust subset, which serves as the
concrete semantics for our abstract interpretation.
It turns the abstract domain and interpretation into dynamic counterparts that record precise aliasing (borrowing) information. We prove that borrow-aware semantics imposes a stack-like ordering among aliased places. The stacked AXM discipline is then formally defined as the absence of out-of-order memory accesses according to this stack ordering. We further show that if stacked AXM is respected, then invalid reference access could not happen, thereby providing a foundation for proving Polonius correct.

\item \textbf{Correctness Proof for Polonius.}
We establish the soundness of abstract interpretation with respect to borrow-aware semantics.
This soundness proof is based on a novel observation that the
type-based pointer analysis used by Polonius can be concretized as
precise aliasing information over a well-typed memory structure. This
correspondence also supports modular analysis for function calls.
By the soundness of
abstract interpretation, we prove that any program admitted by
Polonius respects stacked AXM in borrow-aware semantics,
and is therefore memory safe.

\item \textbf{Rocq Formalization.}
Using CompCert's abstract interpretation
  framework~\cite{Leroy-backend,compcert-va}, we formalize Polonius as
  an instance of Kildall's algorithm. We test this implementation
  extensively to evaluate its agreement with the official Polonius. We
  have also formalized the borrow-aware semantics and correctness
  proof which is generated by LLM based on around 100 pages of
  paper proofs. The complete Rocq formalization is found in the
  supplementary materials.

\end{itemize}

\subsection{Scope of the Paper}

Borrow checking is not the only compiler pass responsible for memory safety. \emph{Type checking} ensures that values in memory are well-typed, so that references could not be invalidated by destructive type casting. \emph{Initialization analysis} ensures no access to uninitialized variables. Furthermore, \emph{move checking} prohibits access to moved in-memory values. For example, the following code is rejected:
\begin{minted}[xleftmargin=1.5em,mathescape=true,escapeinside=||]{rust}
let mut v = Box::new(1);
let mut w = v;  // v no longer owns value
println!("{}", *v); // error: accessing moved memory
\end{minted}
We assume that the above checks have already been performed. Borrow checking is only responsible for ensuring that memory errors could not happen because of accessing an invalid reference \mcode{p}.
Since we assume only well-typed and initialized values are referenced, such an error could happen only if \mcode{p} is a dangling pointer to deallocated memory.
Although this problem seems straightforward, solving it is highly non-trivial.  
Finally, we focus on safe Rust. 
Our work formalized AXM as an invariant which could also work with unsafe Rust; this is left for future work.

 \section{Overview}
\label{sec:overview}

\begin{figure}[htbp]
\definecolor{polredfill}{HTML}{F8CECC}
	\definecolor{polredline}{HTML}{B85450}
	\definecolor{polgreenfill}{HTML}{D5E8D4}
	\definecolor{polgreenline}{HTML}{82B366}
	\centering
	\resizebox{\linewidth}{!}{\begin{tikzpicture}[
			font=\normalfont,
			panel/.style={
				draw,
				line width=0.8pt,
				rounded corners=5pt,
				minimum width=5.25cm,
				minimum height=2.5cm,
				inner xsep=0.20cm,
				inner ysep=0.15cm,
				outer sep=0pt
			},
			annotation/.style={
				font=\normalfont,
				align=center,
				inner sep=2pt
			},
			flow/.style={
				line width=1.1pt,
				-{Stealth[length=2.2mm,width=1.7mm]}
			}
			]
			
\node[panel,fill=polredfill,draw=polredline]
			(constraints) at (0,0) {
				\begin{tabular}{@{}l@{\hspace{0.45em}}l@{}}
					\multicolumn{2}{c}{\large\bfseries Rule-based Polonius (\secref{sec:constraint})} \\[0.15cm]
					$\bullet$ & Domain: regions to sets of loans \\[2pt]
					$\bullet$ & Constraint generation rules\\[2pt]
					$\bullet$ & Constraint solving rules
				\end{tabular}
			};
			\node[panel,fill=polgreenfill,draw=polgreenline,
			right=0.35cm of constraints]
			(dataflow) {
				\begin{tabular}{@{}l@{\hspace{0.45em}}l@{}}
					\multicolumn{2}{c}{\large\bfseries Abstract interpreter for Polonius (\secref{sec:abstract-semantics})} \\[0.15cm]
					$\bullet$ & Domain: regions to region states \\[2pt]
					$\bullet$ & Dataflow operations \\[2pt]
					$\bullet$ & Dataflow solving (Kildall)
				\end{tabular}
			};
			\node[panel,fill=polgreenfill,draw=polgreenline,
			minimum width=5.95cm,right=0.35cm of dataflow]
			(concrete) {
				\begin{tabular}{@{}l@{\hspace{0.45em}}l@{}}
					\multicolumn{2}{c}{\large\bfseries Borrow-aware semantics (\secref{sec:concrete-semantics})} \\[0.15cm]
					$\bullet$ & Domain: value-based memory \& \\
					& references with borrow sets \\[2pt]
					$\bullet$ & Dynamic borrow analysis \& checking
				\end{tabular}
			};
			
\node[annotation] (nll-rfc) at ($(constraints.west)+(-1.12,0.72)$)
			{NLL RFC};
			\node[annotation] at ($(constraints.west)+(-1.12,-0.52)$)
			{Polonius\\(Datalog\\and \mcode{rustc})};
			\draw[decorate,decoration={brace,amplitude=4.5pt},line width=0.9pt]
			($(constraints.north west)+(-0.28,0)$) --
			($(constraints.south west)+(-0.28,0)$);
			
\coordinate (properties-left) at
			($(concrete.east)+(constraints.west)-(nll-rfc.east |- constraints.west)$);
			\node[annotation,anchor=west] (temporal)
			at ($(properties-left)+(0,0.85)$)
			{Temporal\\memory safety};
			\node[annotation,anchor=west] (stacked)
			at ($(properties-left)+(0,-0.85)$)
			{Stacked\\AXM discipline};
			\node[fit=(temporal)(stacked),inner sep=0pt,outer sep=0pt]
			(properties) {};
			\draw[decorate,decoration={brace,amplitude=4.5pt},line width=0.9pt]
			($(concrete.south east)+(0.28,0)$) --
			($(concrete.north east)+(0.28,0)$);
			
\draw[flow] (properties.center |- stacked.north)
			-- node[annotation,right=3pt] (safetyref) {\secref{ssec:memory-safety}}
			(properties.center |- temporal.south);
			
\coordinate (fstart) at
			($(constraints.south west)!0.75!(constraints.south east)$);
			\coordinate (fend) at
			($(dataflow.south west)!0.25!(dataflow.south east)$);
			\draw[flow]
			(fstart) -- ++(0,-0.60) coordinate (fbend)
			-- (fend |- fbend) -- (fend);
			\node[annotation,fill=white,below=2pt]
			at ($(fstart)!0.5!(fend)+(0,-0.60)$)
			{Formalized as};
			
\coordinate (sstart) at
			($(dataflow.south west)!0.75!(dataflow.south east)$);
			\coordinate (send) at
			($(concrete.south west)!0.25!(concrete.south east)$);
			\coordinate (smid) at
			($(sstart)!0.5!(send)+(0,-0.60)$);
			\draw[flow]
			(sstart) -- ++(0,-0.60) coordinate (sbend)
			-- (send |- sbend) -- (send);
			\node[annotation,above=2pt] (soundnessref) at (smid)
			{\secref{ssec:soundness}};
			\node[annotation,fill=white,below=2pt] (soundness) at (smid)
			{Soundness via abstract interpretation};
			
\coordinate (checkstart) at ($(soundness.east)+(0.18,0)$);
			\draw[flow]
			(checkstart)
			-- node[annotation,above=3pt] (successful)
			{Under successful borrow checking}
			(properties.center |- checkstart)
			-- node[annotation,right=3pt] (checkref) {\secref{ssec:stacked-axm}}
			(properties.center |- stacked.south);
			
\node[anchor=south west,inner sep=0pt,outer sep=0pt]
			(rocq) at ($(dataflow.north west)+(0,0.16)$)
			{\includegraphics[width=1.89cm]{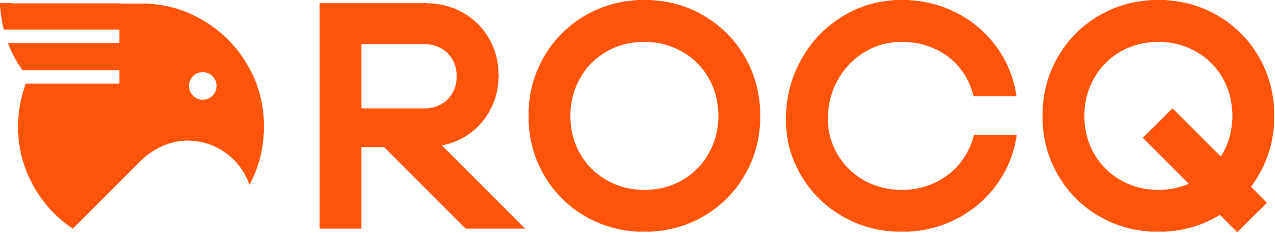}};
			\node[
			draw=black,line width=0.8pt,
			dash pattern=on 3pt off 2.5pt,
			fit=(dataflow)(concrete)(properties)(temporal)(stacked)
			(rocq)(soundness)(successful)(soundnessref)(checkref)(safetyref),
			inner xsep=0.2cm,inner ysep=0.18cm
			] {};
			
		\end{tikzpicture}}
	\caption{Overview of our formalization and verification of Polonius}
	\label{fig:borrow-check-overview}
\end{figure}
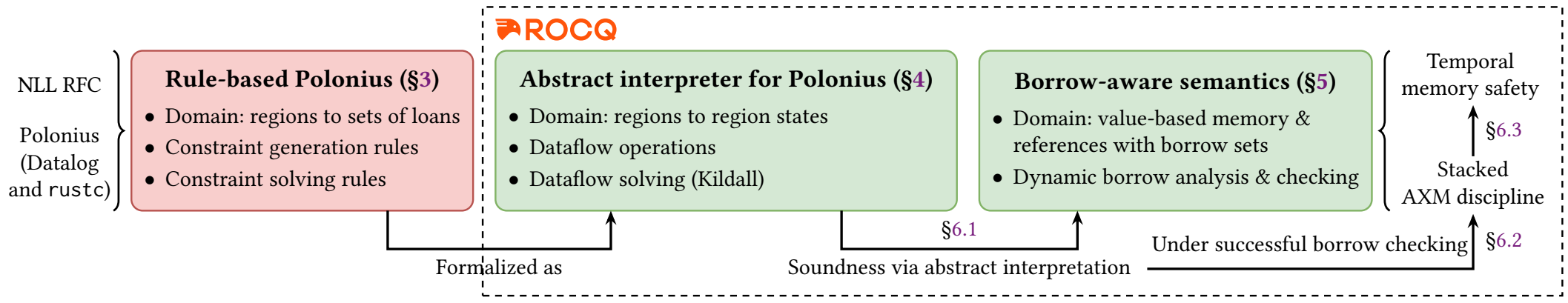

The workflow of this project is depicted in~\figref{fig:borrow-check-overview}. We give a high-level introduction to its components in this section and defer their in-depth discussions to the remaining sections.

\myparagraph{Rule-based Polonius}
This project begins with an investigation of the design and implementation of NLL and Polonius, including the NLL RFC~\cite{rust-nll}, the Datalog rules of Polonius~\cite{rust-polonius,Polonius-thesis}, and the implementation of Polonius in \mcode{rustc}~\cite{rustc-polonius}.
Both NLL and Polonius consist of three steps: generation of borrow constraints, solving of these constraints and detection of borrow errors.
Our first contribution is to reformulate the above algorithmic procedures for Polonius into a \emph{rule-based system}. That is, the constraint generation, solving and borrow checking are all described as a set of derivation rules, thereby eliminating the ad-hocness of the current implementation of Polonius and making it more amenable to formalization.  
We shall discuss those in~\secref{sec:constraint}.

\myparagraph{Abstract Interpreter for Polonius}
Polonius relies on both forward and backward data-flow analysis, where the former propagates borrow constraints and the latter computes liveness information.
This mixed interaction makes it difficult to formalize the derivation rules of Polonius in a theorem prover.  
By replacing the eager borrow checking with lazy checking rules, we obtain a rule-based Polonius without liveness analysis and with only forward data flows.
Then, we formalize Polonius as an abstract interpreter in the form of a \emph{dataflow-based pointer analysis}.
Viewing it as a pointer analysis suggests an abstract domain that maps abstract pointer variables (i.e., regions) to a set of places being borrowed by the pointer (reference).
To deal with complex language features of Rust, we need to design highly nontrivial \emph{abstract domains} and \emph{transfer functions} (i.e., small-step transitions of Rust expressions and statements on abstract domains).
We discuss those in~\secref{sec:abstract-semantics}.

\myparagraph{Borrow-aware Semantics}
To make \emph{static} checking of adherence to AXM possible, Polonius tracks \emph{approximated} information for abstract pointers.
To prove memory safety for concrete execution, we need to accurately capture AXM in the concrete semantics which must track a \emph{precise} set of borrowed places for concrete pointers. 
This results in our borrow-aware semantics.
To avoid reasoning about low-level memory operations, borrow-aware semantics adopts a \emph{value-based memory model} (similar to Low-Level Borrow Calculus, or LLBC~\cite{Aeneas}).
Memory values must be well-typed and memory locations are denoted by named places rather than low-level entities such as blocks of bytes~\cite{compcert-mem-v2}.
Ownership transfer in Rust is modeled by movement of values.
With those features, borrow-aware semantics effectively performs borrow checking \emph{dynamically} using precise aliasing information in the \emph{borrow sets} attached to concrete references.
We discuss those in~\secref{sec:concrete-semantics}.

\myparagraph{Correctness of Polonius}
We first prove that the abstract interpreter for Polonius is sound w.r.t. borrow-aware semantics by establishing a \emph{backward simulation} between them, as depicted in~\figref{fig:soundness-sim} (see~\secref{ssec:soundness}).
By this simulation, successful borrow checking (no borrow error detected by the abstract interpreter) implies no borrow error in the borrow-aware semantics.
We then prove that the stacked AXM discipline is enforced if there is no borrow error in the borrow-aware semantics (see \secref{ssec:stacked-axm}). 
Roughly speaking, the following \emph{borrow chain} is maintained by the borrow-aware semantics:
\[
*p_n \leftarrow \ldots \leftarrow *p_2 \leftarrow *p_1 \leftarrow v
\]
if $p_1$ first borrows $v$, then $p_2$ reborrows $v$ through $p_1$, and so on. We prove that, if there is no borrow error, then Stacked AXM is respected, i.e., that the places on this chain must be accessed \emph{from left to right} (reverse of the borrowing order).
Finally, we prove that the stacked AXM implies memory safety for reference accesses (see~\secref{ssec:memory-safety}). The proof relies critically on an \emph{adequacy} property for borrow-aware semantics: for any two aliased places $p_1$ and $p_2$, if a prefix of $p_1$ could possibly deallocate the memory location that they both reach, then $p_2 \leftarrow p_1$ must be present in the borrow chain. Because stacked AXM does not allow $p_1$ to be used before the last use of $p_2$ and only modification through $p_1$ could invalidate (deallocate) $p_2$, it is impossible to  access dangling references through $p_2$. Therefore, every dereferencing operation is memory safe for borrow checked programs.

\begin{figure}[t]
	\begin{subfigure}{0.23\textwidth}
		\centering
		\begin{tikzpicture}
			\node (s1) {$S_1$};
			\node[right = 1 of s1] (s2) {$S_2$};
			\node[below = 0.8 of s1] (sp1) {$S_1'$};
			\node (sp2) at (sp1 -| s2) {$S_2'$};
			
			\draw[-latex,dashed] (s1) -- (s2);
			\draw[-latex] (sp1) -- (sp2);
			\draw[<->] (s1) -- node[left] {$\gamma$} (sp1);
			\draw[<->,dashed] (s2) -- node[left] {$\gamma$} (sp2);
		\end{tikzpicture}
		\caption{Internal simulation}
	\end{subfigure}\hfill \begin{subfigure}{0.44\textwidth}
		\centering
		\begin{tikzpicture}
			\node (s1) {$S_1$};
			\node[right = 2 of s1] (s2) {$S_2$};
			\node[below = 0.8 of s1] (sp1) {$S_1'$};
			\node (sp2) at (sp1 -| s2) {$S_2'$};
			\node (steps) at ($(sp1)!0.5!(sp2)$) {$\mydots$};
			
			\draw[-latex,dashed] (s1) -- (s2);
			\draw[-latex] (sp1) -- node[above,font=\scriptsize] {call} (steps);
			\draw[-latex] (steps) -- node[above,font=\scriptsize] {return} (sp2);
			\draw[<->] (s1) -- node[left] {$\gamma$} (sp1);
			\draw[<->,dashed] (s2) -- node[left] {$\gamma$} (sp2);

\end{tikzpicture}
		\caption{Simulation of function call and return}
		\label{sfig:soundness-call-sim}

\end{subfigure}\hfill \begin{subfigure}{0.31\textwidth}
		\centering
		\begin{tikzpicture}
			\node (s1) {$S_1$};
			\node[right = 1 of s1] (f1) {\small Borrow Error};
			\node[below = 0.8 of s1] (sp1) {$S_1'$};
			\node (f2) at (sp1 -| f1) {\small Borrow Error};
			
			\draw[-latex,dashed] (s1) -- (f1);
			\draw[-latex] (sp1) -- (f2);
			\draw[<->] (s1) -- node[left] {$\gamma$} (sp1);
		\end{tikzpicture}
		\caption{Borrow-error simulation}
	\end{subfigure}
	\begin{flushleft} {\footnotesize The top row denotes transitions over
			abstract states $S_1$ and $S_2$ via the transfer function of the
			abstract interpreter for Polonius. The bottom row denotes
			transitions over concrete states $S_1'$ and $S_2'$ in the
			borrow-aware semantics. An invariant $\gamma$ is maintained by internal
			steps throughout the entire execution. For a function call, the
			entire concrete execution from call to return must be simulated by
			a single transfer step at the abstract interpreter, since Polonius
			performs modular analysis. When the borrow-aware semantics detects
			a borrow error, so does the abstract interpreter.}
\end{flushleft}
	
	\caption{Backward simulation between abstract interpreter and borrow-aware semantics}
	\label{fig:soundness-sim}
\end{figure}

\begingroup
\section{Language and Rule-based Polonius}\label{sec:constraint}

\subsection{A Core language for Intermediate Representation of Rust}

\begin{figure}[ht]
  \smaller
  \arraycolsep=1pt
  \centering
\[
\begin{array}{ll} 
\begin{array}[t]{lrll}
& x, \id, \fid, n \in \app & \mathbb{N} & \text{identifiers} \\
    
& \rho, r \in R ::=\app & \mathbb{N} & \text{regions} \\
    
& q        ::=\app && \text{projections} \\
&                  & [.] & \text{base cases} \\
&                  & \deref^b q  & \text{Box dereferences} \\
&                  & \deref^r q  & \text{reference dereferences} \\
&                  & q.\fid & \text{field selections}\\
  
& p \in P ::=\app & q[x] & \text{places}\\

& \omega \in \Omega ::=\app & \texttt{mut} \stk \texttt{shr} & \text{mutability} \\

& \op ::=\app && \text{operands} \\
&                 &  \texttt{true} \stk \texttt{false} \stk \mathit{n_{\kwd{i32}}}\dots & \text{constants} \\
&                 & \copyp{p} & \text{scalar copies} \\
&                 & \move{p} & \text{ownership transfers}\\

& e \in E ::=\app && \text{expressions (r-values)} \\
&                 & \op & \text{operands}\\    
&                 & \&r \app \omega\app p & \text{borrow expressions} \\
&                 & \texttt{uop}\app \op \stk \op_1 \app\texttt{bop}\app \op_2 & \text{operators} \\
&                 & \texttt{discr}(p) & \text{enum discriminants}\\    
&                 & \texttt{Box}(\op) & \text{heap-allocated values}\\
&                 & \enumrv{\op} & \text{enum constructors}\\
\\
& \prog ::=\app & \overrightarrow{d}& \text{programs}
\end{array}
\begin{array}[t]{lrll}

& i \in I ::= \app && \text{instructions} \\
&                  & \texttt{goto}(n) & \text{unconditional jumps}\\
&                 & (\sassign{p}{e}, n) & \text{assignments}\\

&                  & (p:=id_{f}(\overrightarrow{\op}),n) & \text{function calls}\\
&                 & (\drop{p},n) & \text{resource deallocations}\\

&                  &\texttt{cond}(\op, n_1, n_2) & \text{conditional jumps}\\
&                  &\texttt{return}(\retx) & \text{function returns}\\
& g\in G ::= \app & \{\texttt{entry}: \mathbb{N}, & \\
&                 &  \app\app \texttt{body}: \mathbb{N} \rightarrowpar I\} & \text{control-flow graphs} \\

& \tau \in \Tau ::=\app && \text{types} \\
&                 & \texttt{bool} \stk \texttt{i32} \stk \dots & \text{base types}\\
&                 & \tbox{\tau} & \text{heap pointers} \\
&                 & \tref[r]{\tau} & \text{references}\\
&                 & \tstructregs{\id}{\overrightarrow{\rho}} & \text{structs}\\
&                 & \tenumregs{\id}{\overrightarrow{\rho}} & \text{enums} \\
&                 & \ensuremath{\texttt{fn}\app\langle \overrightarrow{\rho}, \overrightarrow{\rsubset{\rho_1}{\rho_2}}\rangle(\targs)\rightarrow\tau} & \text{function types} \\

& \mathit{d} \in \mathit{D}::=\app && \text{top-level definitions} \\
&                                  & \texttt{fn}\app \id_f\langle \overrightarrow{\rho}, \overrightarrow{\rsubset{\rho_1}{\rho_2}}\rangle&  \\
&                 &(\overrightarrow{x_{\text{param}}:\tau})\rightarrow \tau  & \\
&                 & \app = \{(\overrightarrow{x_{\text{local}}:\tau}), \retx:\tau , g\} & \text{function definitions} \\
&                 & \tstructregs{\id}{\overrightarrow{\rho}}(\overrightarrow{\fid:\tau}) & \text{struct definitions} \\
&                 & \tenumregs{\id}{\overrightarrow{\rho}}(\overrightarrow{\fid:\tau}) & \text{enum definitions}
\end{array}
\end{array}
\]
\caption{Syntax of Core MIR}
\label{fig:core-lang}
\end{figure}

Our borrow checker operates on a core subset of Rust's Mid-level
Intermediate Representation (MIR)~\cite{mir-rfc}, defined
in~\figref{fig:core-lang}, which we call \emph{Core MIR}.
We retain features central to
our analysis, including references, \mcode{Box} pointers,
composite data types (structs and enums), and function calls,
while excluding slices, type polymorphism, traits, closures, and unsafe features such as raw pointers.

A program consists of three kinds of top-level definitions: functions,
structs, and enums.  Each definition declares a list of \emph{region
  parameters} $\overrightarrow{\rho}$, corresponding to lifetime
parameters in the NLL terminology.  Regions appearing in types of
function parameters or types of struct and enum fields must come from
these parameters.  Function signatures also declare a list of region
relations $\overrightarrow{\rsubset{\rho_1}{\rho_2}}$, whose role in
borrow checking is explained later.
Each function declares all its local variables and their types at the
entry of the function.  A special variable $\retx$ holds the return
value, which is returned by the instruction $\texttt{return}(\retx)$.
The function body $g$ is a control-flow graph (CFG), consisting of an
entry node and a partial map from node labels (denoted by $n$) to
instructions. We shall write $\succ(n)$ for the set of successor
nodes of $n$.
In \mcode{rustc}, a control-flow block is a sequence of
statements followed by a terminator. 
We instead place a single
instruction at each CFG node.

We follow MIR's design of expressions (i.e., r-values) and
operands~\cite{mir-rfc}.  MIR does not contain nested expressions.
Source-level nested expressions are lowered into assignments to
temporary variables, followed by their uses in the operands.
The
expression $\texttt{Box}(\op)$ allocates heap memory to store the
value of $\op$. $\enumrv{\op}$ constructs an enum value using
the $\fid$ constructor of the enum type named $\id_{\texttt{enum}}$.
Operands explicitly distinguish whether a place is used by move or by copy,
allowing the analysis to track their different effects on ownership.
Places identify memory locations (i.e., l-values).  A place $p$ consists of a variable
with a sequence of projections applied to it.  A projection selects a
field of a struct or enum, or dereferences a \mcode{Box} or reference,
where the annotations $b$ and $r$ can be omitted when we do not need
to distinguish them.

A borrow expression $\&r \app \omega\app p$ creates a borrow of place
$p$ with mutability $\omega$ (\mcode{mut} for mutable or \mcode{shr} for shared, we also write them as \mcode{m} and \mcode{s}, respectively).  
Note that $r$ is a \emph{temporary region} alive only in this very expression. Despite that, it plays the critical role of collecting loans from borrows and reborrows in borrow checking, as we shall see later.
Regions introduced for borrow expressions at different program points
are distinct from one another and from the regions appearing in
variable and parameter types.  If a place has reference type, its type
is also annotated with a region, e.g., $\tref[r][\omega]{\tau}$.
We use $r$ as the default notation for regions (e.g., regions in the types of local variables). We use $\rho$ specifically for region parameters
declared by functions, structs, and enums, and for their occurrences
in the types of function parameters.

\subsection{A Rule-based Formulation of Polonius Borrow
  Checking}\label{ssec:polonius-constraint}

\begin{figure}[t]
  \small
  \begin{subfigure}[t]{\textwidth}
  \begin{mathpar}
    \inferrule[Pol-Sub-Base]{
      \tau = \mcode{i32}, \mcode{u32}, ...
    }{
      \tau \subty \tau \rightsquigarrow \emptyset
    }
\quad          
    \inferrule[Pol-Sub-Mut]{
      \tau_1 \subty \tau_2 \rightsquigarrow C_1\quad
      \tau_2 \subty \tau_1 \rightsquigarrow C_2
    }{
      \tref[r_1][\texttt{mut}]{\tau_1} \subty \tref[r_2][\texttt{mut}]{\tau_2} \rightsquigarrow
      \rawset{r_1 \subseteq r_2} \cup C_1 \cup C_2
    }
\quad
    \inferrule[Pol-Sub-Shr]{
      \tau_1 \subty \tau_2 \rightsquigarrow C
    }{
      \trefs{r_1}{\tau_1} \subty \trefs{r_2}{\tau_2} \rightsquigarrow \rawset{r_1 \subseteq r_2} \cup C
    }

    \inferrule[Pol-Bor-Expr]{
      p_2 = \tref[r][\omega]{p_1} \textbf{ at } n\quad
      C_1 = \rawset{p_1^{\omega}} \subseteq r \quad
      r_1 \subseteq r_2 \in C_1 \cup C_2\\\\
      C_2 = \pset{r'\subseteq r}{\forall (p': \&\,r'\, \_ \, \_), \app \prefix( \deref p', p_1)}
    }{
      r_1 \subseteq r_2 \textbf{ at } n
    }

    \inferrule[Pol-Assign]{
      p_2: \tau_2 = \tref[r][\omega]{p_1} \textbf{ at } n\quad
      \tau_1 = \typeof{p_1} \\\\
      \tref[r][\omega]{\tau_1}  \subty \tau_2 \rightsquigarrow C\quad
      r_1 \subseteq r_2 \in C
    }{
      r_1 \subseteq r_2 \textbf{ at } n
    }
  \end{mathpar}
  \caption{The generation rules of subset constraints for assignment}
  \label{sfig:borrow-constraint}
\end{subfigure}

  \begin{subfigure}[t]{\textwidth}
  \begin{mathpar}
    \inferrule[Pol-Subset-Flow]{
      r_1 \subseteq r_2 \textbf{ at } n\\\\
      r_1\app r_2 \textbf{ live at } \succ(n)
    }{
      r_1 \subseteq r_2 \textbf{ at } \succ(n)
    }

    \inferrule[Pol-Loan-Subset]{
      \rawset{p^{\omega}} \subseteq r_1 \textbf{ at } n \\\\
      r_1 \subseteq r_2 \textbf{ at } n
    }{
      \rawset{p^{\omega}} \subseteq r_2 \textbf{ at } n
    }

    \inferrule[Pol-Loan-Flow]{
      \rawset{p^{\omega}} \subseteq r \textbf{ at } n \quad
      r \textbf{ live at } \succ(n)\\\\
      p^{\omega} \textbf{ not killed at } n
    }{
      \rawset{p^{\omega}} \subseteq r \textbf{ at } \succ(n)
    }

    \inferrule[Pol-Loan-Live]{
      \rawset{p^{\omega}} \subseteq r \textbf{ at } n\\\\
      r \textbf{ live at } n
    }{
      p^{\omega} \textbf{ live at } n
    }
  \end{mathpar}
\caption{The inference rules for loan analysis}
  \label{sfig:loan-analysis}
\end{subfigure}

  \begin{subfigure}[t]{\textwidth}
  \begin{mathpar}
    \inferrule[Pol-Write-Invalid]{
     p' \textbf{ written at } n\quad
     p \# p'
    }{
      p^{\omega} \textbf{ invalidated at } n
    }
    
    \inferrule[Pol-Read-Invalid]{
     p' \textbf{ read at } n\quad
     p \# p'
    }{
      p^{\texttt{m}} \textbf{ invalidated at } n
    }
    
    \inferrule[Pol-Borrow-Check]{
      p^{\omega} \textbf{ invalidated at } n\quad
      p^{\omega} \textbf{ live at } n
    }{
      \textbf{borrow-error at } n
    }
  \end{mathpar}
\caption{The inference rules for error detection}
  \label{sfig:error-detection}
\end{subfigure}
\caption{The core inference rules for the Polonius borrow checker}
\label{fig:polonius-inferences}
\begin{center}
  \smaller
\begin{tikzpicture}[
  pol node/.style={
    draw,
    align=center,
    minimum width=2.5cm,
    minimum height=0.8cm
  },
  pol arrow/.style={-Stealth, thick}
]
  \def\polBlockSep{1cm}
  \def\polForkSep{0.5cm}

  \node (mir) {MIR};
  \node[pol node, above right=-0.1cm and \polBlockSep of mir] (subset)
    {Subset constraints\\(\figref{sfig:borrow-constraint})};
  \node[pol node, below right=-0.1cm and \polBlockSep of mir] (live)
    {Liveness constraints\\(for ``$r \textbf{ live at } n$'')};
  \node[pol node, right=\polBlockSep of $(subset.east)!0.5!(live.east)$] (analysis)
    {Loan analysis (\figref{sfig:loan-analysis}) \\(constraint solving)};
  \node[pol node, right=\polForkSep of analysis] (detect)
    {Detect borrow errors (\figref{sfig:error-detection})\\and illegal region relations};
  \node[right=0.55cm of detect] (ok) {OK/Error};

  \coordinate[right=\polForkSep of mir] (fork);
  \coordinate[right=\polForkSep of $(subset.east)!0.5!(live.east)$] (join);

  \draw[thick] (mir.east) -- (fork);
  \draw[pol arrow] (fork) |- (subset.west);
  \draw[pol arrow] (fork) |- (live.west);
  \draw[thick] (subset.east) -| (join);
  \draw[thick] (live.east) -| (join);
  \draw[pol arrow] (join) -- (analysis.west);
  \draw[pol arrow] (analysis.east) -- (detect.west);
  \draw[pol arrow] (detect.east) -- (ok.west);
\end{tikzpicture}
\end{center}
\caption{Structure of the Polonius borrow checker}
\label{fig:polonius-borrow-check}
\end{figure}

In the Rust compiler, borrow checking is implemented as a highly
engineered and optimized system. Nevertheless, it roughly consists of three steps: generating region constraints to
capture the borrowing information \emph{generated} at each program
point, solving these constraints to \emph{propagate} this information
across program points, and detecting errors, i.e., violations of the
AXM discipline, based on the solution.
Figure~\ref{fig:polonius-borrow-check} illustrates the three steps of
Polonius. Correspondingly, we present the core inference rules used in
these three steps in \figref{fig:polonius-inferences}. The
presentation style is inspired
by~\cite{will-conceptual}, but we give a more formal 
account that covers the whole process from constraint generation to
error detection.
Polonius mainly relies on two kinds of constraints: \emph{subset
  constraints} and \emph{liveness constraints}.
Assignment is the most important case for understanding how subset
constraints are generated, since creating a new borrow through borrow
expressions can only appear on the right-hand side of assignments. We
present the constraint-generation rules for assignments
in~\figref{sfig:borrow-constraint}. They summarize the reborrow
and subtyping rules introduced in the NLL
RFC~\cite{rust-nll,rust-subtyping}, but are presented in the style of
Polonius, using subset relations ($r_1\subseteq r_2$) between regions
rather than outlives relations ($r_1:r_2$) between lifetimes as in
NLL.
The rules in~\figref{sfig:loan-analysis}
and~\figref{sfig:error-detection} follow the Datalog rules of
Polonius~\cite{rust-polonius}. These rules rely on the subset
constraints and liveness constraints generated
from~\figref{sfig:borrow-constraint}, mirroring how the Datalog rules
rely on "input facts" generated from \mcode{rustc}.

Reading the rules from bottom to top, \Rule{Pol-Borrow-Check} reports
a borrow-checking error when a loan is invalidated while it is still
alive, i.e., \emph{AXM is violated}.
A \emph{loan} takes the form $p^\omega$, denoting that the place $p$
is borrowed with mutability $\omega$.
That is, loans represent the borrowing information computed by the borrow
checkers.
Intuitively, a loan $p^\omega$ is alive means that a borrower of
$p$ is still alive, so $p$ cannot be accessed in a way that conflicts
with this borrow.
The \textbf{invalidated at} judgment is defined by two cases.
\Rule{Pol-Write-Invalid} states that a loan on a place $p$ is
invalidated by a write to any place $p'$ that overlaps with $p$
(denoted as $p \# p'$). Here, $p$ and $p'$ overlap if one is a \emph{prefix}
of the other. By definition, $p$ is a prefix of $p'$ iff
$\exists q,\app p'= q[p]$.
\Rule{Pol-Read-Invalid} states that a mutable loan on a
place $p$ is invalidated by a read of its overlapped place. A place
$p$ is considered written if it is assigned, dropped, moved, or
borrowed mutably. Otherwise, if $p$ is copied or borrowed sharedly, it
is considered read.
The judgment ``$p^\omega\app\textbf{live at}\app n$'' is derived by
\Rule{Pol-Loan-Live} as the result of loan analysis. Loan analysis
computes which loans are contained in each region at each program
point, i.e., judgments of the form
``$\rawset{p^\omega} \subseteq r \app\textbf{at}\app n$''.
A loan is alive when it belongs to a live region, written as
``$r \textbf{ live at } n$'', meaning that $r$ is alive at the start
of $n$.  Such constraints are computed by a standard liveness analysis 
of which we omit a discussion.
In the following paragraphs, we discuss how the rules in~\figref{fig:polonius-inferences} collaborate to perform borrow checking as realized in Polonius.

\myparagraph{Borrows and reborrows}
We begin with a discussion of how borrow checking is performed in presence of references and reborrowing of values through references.
Consider the following example which would cause a memory error
without borrow checking. We explain how Polonius rejects this program
using the rules in~\figref{fig:polonius-inferences}.  We present the
example in Rust surface syntax for readability, with line numbers
serving as CFG nodes. The comments show the
subset constraints generated at this node (by~\figref{sfig:borrow-constraint}) and the computed live loans
at the next node (by~\figref{sfig:loan-analysis}).
\begin{minted}[xleftmargin=1.5em,formatcom=\RustCodeLineNumbers,mathescape=true,escapeinside=||]{rust}
let mut v = Box::new(1); // 'r1 and 'r2 are temporary regions and would be dead immediately
let mut x: &'x mut Box<i32> = &'r1 mut v; // $\rawset{\mcode{v}^{\mutx}}\subseteq$'r1 (by $\textup{\Rule{Pol-Bor-Expr}}$)
// 'r1$\subseteq$'x (by $\textup{\Rule{Pol-Sub-Mut}}$ and $\textup{\Rule{Pol-Assign}}$)
// Live loans at $l_{3}$: $\rawset{\mcode{v}^{\mutx}}\subseteq$'x (by $\textup{\figref{sfig:loan-analysis}}$); 'x is alive until $l_{3}$
let mut y: &'y i32 = &'r2 **x; // $\rawset{\mcode{\deref\deref x}^{\texttt{s}}}\subseteq$'r2, 'x$\subseteq$'r2 (by $\textup{\Rule{Pol-Bor-Expr}}$)
// 'r2$\subseteq$'y (by $\textup{\Rule{Pol-Sub-Shr}}$ and $\textup{\Rule{Pol-Assign}}$)
// Live loans at $l_{4}$: $\rawset{\mcode{v}^{\mutx}, \mcode{\deref\deref x}^{\texttt{s}}}\subseteq$'y (by $\textup{\figref{sfig:loan-analysis}}$); 'y is alive until $l_{5}$
v = Box::new(2); // borrow-error by $\textup{\Rule{Pol-Write-Invalid}}$ and $\textup{\Rule{Pol-Borrow-Check}}$
println!("{}", *y); // Without borrow checking, this would access freed memory.
\end{minted}
In the example, the final use of \mcode{y} keeps \ticktt{y} alive
across the reassignment of \mcode{v}.  Since \ticktt{y} contains
$\loan{v}{m}$, this mutable loan is still alive when \mcode{v} is
written\footnote{After lowering to our Core MIR, the source-level
  assignment \mcode{v = Box::new(2)} is expanded into \mcode{drop(v);
    v = Box::new(2)}, so that the old value is first released before
  the new value is assigned. Dropping counts as a
  write and incurs the error.}.  The checker therefore rejects the
reassignment using the rules in~\figref{sfig:error-detection}.  This
prevents the later access through \mcode{y} from reaching memory that
would have been freed by replacing \mcode{v} with a new \mcode{Box}
value.

To understand why \ticktt{y} contains $\loan{v}{m}$, we examine
the subset constraints.  As in constraint-based pointer
analysis~\cite{andersen1994program}, these constraints propagate
points-to information between regions.
A basic case is issuing a loan for a borrow assignment of the form
$p_2:\tref[r_1][\omega]{\tau} = \tref[r_2][\omega]{p_1}$. $C_1$ in
\Rule{Pol-Bor-Expr} records the directly issued loan through
$\rawset{p_1^\omega}\subseteq r_2$.  \Rule{Pol-Assign} then generates
constraints through subtyping, including $r_2\subseteq r_1$, to
propagate the loan to the assignee $p_2$.
In the example, the assignments to \mcode{x} and \mcode{y} introduce
$\loan{v}{m}$ into \ticktt{x} and $\loan{\deref\deref x}{s}$ into
\ticktt{y}, respectively.  

The above constraints alone do not explain how
$\loan{v}{m}$ flows into \ticktt{y}.
This propagation relies on the \emph{reborrow constraints} in $C_2$,
the second part of \Rule{Pol-Bor-Expr}.
Intuitively, if the borrowed place (i.e., $p_1$) already borrows from
other places, $C_2$ collects the loans representing those places so
that they are \textbf{re}borrowed by the assignee.
This is realized by collecting subset constraints on
\emph{prefixes}\footnote{The NLL RFC~\cite{rust-nll} uses
  \emph{supporting prefixes}. Starting from a place, their collection
  removes projections and stops at a dereference of a shared reference.
  We use ordinary prefixes in the main text to simplify the presentation.
  The appendix and our Rocq formalization use supporting prefixes.}
of $p_1$ with reference types w.r.t. $p_2$
as described in $C_2$.
When \mcode{y} is created, $C_2$ generates
$\ticktt{x}\subseteq\ticktt{r2}$.  Together with
$\ticktt{r2}\subseteq\ticktt{y}$ from the assignment, this propagates
$\loan{v}{m}$ from \ticktt{x} through \ticktt{r2} to \ticktt{y}.
The loan is then carried to the next node by \Rule{Pol-Loan-Flow}.
Together with the liveness of \ticktt{y}, this propagation allows the
checker to detect the conflicting write to \mcode{v}.

\myparagraph{Stacked AXM discipline} \Rule{Pol-Bor-Expr} ensures that
a newly created borrow not only creates a loan for its direct lender,
but also inherits all the loans from that lender through reborrowing.
This explains why mutable references to the same location should be
used in a first-created-last-used order, i.e., following the stacked
AXM discipline.  Consider the following example
from~\figref{fig:intro-exm}:
\begin{minted}[xleftmargin=1.5em,formatcom=\RustCodeLineNumbers,mathescape=true,escapeinside=||]{rust}
let mut v = Box::new(1);
let mut x: &'x mut i32 = &mut *v; // $\rawset{\mcode{\deref v}^{\mutx}}\subseteq$'x (by $\textup{\Rule{Pol-Assign}}$)
// Live loans at $l_{3}$: $\rawset{\mcode{\deref v}^{\mutx}}\subseteq$'x (by $\textup{\figref{sfig:loan-analysis}}$); 'x is alive until $l_{4}$
let mut y: &'y mut i32 = &mut *x; // $\rawset{\mcode{\deref x}^{\mutx}}\subseteq$'y, 'x$\subseteq$'y (by $\textup{\Rule{Pol-Assign}}$)
// Live loans at $l_{4}$: $\rawset{\mcode{\deref v}^{\mutx}}\subseteq$'x, $\rawset{\mcode{\deref v}^{\mutx}, \mcode{\deref x}^{\mutx}}\subseteq$'y (by $\textup{\figref{sfig:loan-analysis}}$); 'y is alive until $l_{5}$
*x = 2; // borrow-error by $\textup{\Rule{Pol-Write-Invalid}}$ and $\textup{\Rule{Pol-Borrow-Check}}$
*y = 3;
\end{minted}
Here, \mcode{\deref y}, \mcode{\deref x}, and \mcode{\deref v} reach
the same location and are all mutable, but their loan sets have a
containment relation.  Since \mcode{y} carries a loan on \mcode{\deref
  x} and \mcode{x} carries a loan on \mcode{\deref v}, this containment
relation, together with \Rule{Pol-Borrow-Check}, enforces the
stack order of first using \mcode{y}, then \mcode{x}, and~finally~\mcode{v}.

\myparagraph{Nested references}
Like pointers-to-pointers in C/C++, Rust supports \emph{nested references}. The main challenge is that nested references introduce indirect
updates which require \emph{merging} of regions. That is, through one
reference, the program may modify another reference and thereby change
the places it borrows.  As a result, the loans associated with a
region may change even when that region is not updated directly.  Rust
handles this case through the subtyping rules~\cite{rust-subtyping}
for assignments, which can merge regions appearing at corresponding
levels of nested reference types. Consider an example that may cause a
memory error through an indirect update:
\begingroup
\begin{minted}[xleftmargin=1.5em,formatcom=\RustCodeLineNumbers,mathescape=true,escapeinside=||]{rust}
let mut v1 = 1; let mut v2 = Box::new(2);
let mut x: &'x mut i32 = &'r1 mut v1; // $\rawset{\mcode{v1}^{\mutx}}\subseteq$'r1 (by $\textup{\Rule{Pol-Bor-Expr}}$)
// 'r1$\subseteq$'x (by $\textup{\Rule{Pol-Sub-Mut}}$ and $\textup{\Rule{Pol-Assign}}$)
// Live loans at $l_{3}$: $\rawset{\mcode{v1}^{\mutx}}\subseteq$'x (by $\textup{\figref{sfig:loan-analysis}}$); 'x is alive until $l_{6}$
let mut y: &'y1 mut &'y2 mut i32 = &'r2 mut x; // $\rawset{\mcode{x}^{\mutx}}\subseteq$'r2 (by $\textup{\Rule{Pol-Bor-Expr}}$)
// 'r2$\subseteq$'y1, 'x$\subseteq$'y2, 'y2$\subseteq$'x (by $\textup{\Rule{Pol-Sub-Mut}}$ and $\textup{\Rule{Pol-Assign}}$)
// Live loans at $l_{4}$: $\rawset{\mcode{x}^{\mutx}}\subseteq$'y1, $\rawset{\mcode{v1}^{\mutx}}\subseteq$'y2$\equiv$'x (by $\textup{\figref{sfig:loan-analysis}}$); 'y1 and 'y2 are alive until $l_{4}$
*y = &'r3 mut *v2; // $\rawset{\mcode{\deref v2}^{\mutx}}\subseteq$'r3 (by $\textup{\Rule{Pol-Bor-Expr}}$)
// 'r3$\subseteq$'y2 (by $\textup{\Rule{Pol-Sub-Mut}}$ and $\textup{\Rule{Pol-Assign}}$); 'x$\subseteq$'y2, 'y2$\subseteq$'x (by $\textup{\Rule{Pol-Subset-Flow}}$)
// Live loans at $l_{5}$: $\rawset{\mcode{v1}^{\mutx}, \mcode{\deref v2}^{\mutx}}\subseteq$'x (by $\textup{\figref{sfig:loan-analysis}}$)
v2 = Box::new(3); // borrow-error by $\textup{\Rule{Pol-Write-Invalid}}$ and $\textup{\Rule{Pol-Borrow-Check}}$
*x = 4; // Without borrow checking, this would access freed memory.
\end{minted}
\endgroup
At line~3, creating \mcode{y} makes \mcode{\deref y} and \mcode{x}
denote the same memory location. Polonius records this aliasing fact
by generating the two subset constraints
$\ticktt{x}\subseteq\ticktt{y2}$ and $\ticktt{y2}\subseteq\ticktt{x}$
via \Rule{Pol-Sub-Mut}, i.e., $\ticktt{y2}\equiv\ticktt{x}$.
This equality is crucial at line 4. The assignment of \mcode{\deref y}
also updates the reference value stored in \mcode{x}, so the loan
$\loan{\mcode{\deref v2}}{\mutx}$ flows into \ticktt{y2} and hence also into
\ticktt{x}. As a result, the later write to \mcode{v2} invalidates
$\loan{\mcode{\deref v2}}{\mutx}$ that is still alive as it is contained in
the live region \ticktt{x}, which causes borrow-error.

\myparagraph{References in composite data types}
The preceding examples show how regions can be equated during
analysis. Regions can also be equal by definition.  In a struct/enum,
different reference fields may share the same region parameter and
thus the same loan set.  Consider this example:
\begin{minted}[xleftmargin=1.5em,formatcom=\RustCodeLineNumbers,mathescape=true,escapeinside=||]{rust}
// struct Point<'a> {x: &'a mut i32, y: &'a mut Box<i32>}
let mut v1 = 1; let mut v2 = Box::new(2); let mut p: Point<'p>;
p.x: &'p mut i32 = &'r1 mut v1; // $\rawset{\mcode{v1}^{\mutx}}\subseteq$'r1, 'r1$\subseteq$'p
p.y: &'p mut Box<i32> = &'r2 mut v2; // $\rawset{\mcode{v2}^{\mutx}}\subseteq$'r2, 'r2$\subseteq$'p
// Live loans at $l_{4}$: $\rawset{\mcode{v1}^{\mutx}, \mcode{v2}^{\mutx}}\subseteq$'p; 'p is alive until $l_{4}$
let mut z: &'z mut i32 = &'r3 mut **(p.y); // $\rawset{\mcode{\deref\deref(p.y)}^{\mutx}}\subseteq$'r3, 'p$\subseteq$'r3, 'r3$\subseteq$'z
// Live loans at $l_{5}$: $\rawset{\mcode{v1}^{\mutx}, \mcode{v2}^{\mutx}, \mcode{\deref\deref(p.y)}^{\mutx}}\subseteq$'z; 'z is alive until $l_{6}$
v1 = 4; // borrow-error by $\textup{\Rule{Pol-Write-Invalid}}$ and $\textup{\Rule{Pol-Borrow-Check}}$
*z = 3;
\end{minted}
The assignments to \mcode{p.x} and \mcode{p.y} introduce loans
$\mcode{v1}^\mutx$ and $\mcode{v2}^\mutx$ into \ticktt{p},
respectively.  When \mcode{z} is created, the reborrow constraint
propagates both loans into \ticktt{z}.  As a result, borrow checking
rejects the assignment to \mcode{v1}, even though the last two
statements access disjoint memory locations.
This imprecision arises because a single region merges borrowing
information from multiple reference fields.  Such imprecision is
unavoidable when supporting references in composite data types,
especially recursive types that may contain an unbounded number of
references.

\myparagraph{Kill loans}
For the judgment ``$p^\omega\app\textbf{not killed at}\app n$'' in
\Rule{Pol-Loan-Flow}, informally, if a loan has the form
$\deref^r p^\omega$ and $p$ is reassigned, then $\deref^r p^\omega$ is
considered killed.  Intuitively, $p$ now refers to a new value, so the
previous borrowing relation through $p$ no longer holds.
\begin{minted}[xleftmargin=1.5em,formatcom=\RustCodeLineNumbers,mathescape=true,escapeinside=||]{rust}
let mut v1 = 1; let mut v2 = 2;
let mut x: &'x mut i32 = &'r1 mut v1; // $\rawset{\mcode{v1}^{\mutx}}\subseteq$'r1, 'r1$\subseteq$'x
let y: &'y mut i32 = &'r2 mut *x; // $\rawset{\mcode{\deref x}^{\mutx}}\subseteq$'r2, 'x$\subseteq$'r2, 'r2$\subseteq$'y
// Live loans at $l_{4}$: $\rawset{\mcode{v1}^{\mutx}, \mcode{\deref x}^{\mutx}}\subseteq$'y; 'y is alive until $l_{5}$
x = &'r3 mut v2; // $\rawset{\mcode{v2}^{\mutx}}\subseteq$'r3, 'r3$\subseteq$'x; $\mcode{\deref x}^{\mutx}$ is killed as x is reassigned
// Live loans at $l_{5}$: $\rawset{\mcode{v2}^{\mutx}}\subseteq$'x, $\rawset{\mcode{v1}^{\mutx}}\subseteq$'y; 'x is alive until $l_{5}$
*x = 3; *y = 4; // Both accesses are OK
\end{minted}
At line~4, \mcode{x} is reassigned to point to \mcode{v2}, while
\mcode{y} still points to \mcode{v1}.  The reassignment kills the loan
$\loan{\mcode{\deref x}}{\mutx}$ created at line~3, so it is not
propagated into \ticktt{y} at line~5. 
\footnote{At line~4, applying only \Rule{Pol-Write-Invalid} would
  invalidate the loan $\loan{\mcode{\deref x}}{\mutx}$ and report a borrow
  error because \ticktt{y} is still alive. The actual borrow checker
  distinguishes \emph{shallow} and \emph{deep} accesses~\cite{rust-nll}.
  Reassigning \mcode{x} is a shallow access: it updates the reference
  stored in \mcode{x} without accessing \mcode{\deref x}. Therefore, it
  does not invalidate this loan. The appendix and our Rocq formalization
  include this distinction.}
Thus,
\mcode{x} and \mcode{y} can subsequently be used independently to
modify \mcode{v2} and \mcode{v1}, respectively. If
$\loan{\mcode{\deref x}}{\mutx}$ were not killed, using \mcode{x}
before \mcode{y} would be rejected.  Loan killing is an important
feature to allow the checker to accept such reassignment-then-use
patterns, as discussed in the problem case \#4 of the NLL
RFC~\cite{rust-nll}.

\myparagraph{Control flow, function calls and function returns}
The examples above illustrate the main concepts of borrow checking.
We cover the remaining details when presenting Polonius as an abstract
interpreter in the following sections.  In particular, we define an
abstract domain with join operations to merge states from different
control-flow paths, along with transfer functions for function calls
and returns.
Polonius also detects \emph{illegal region
  relations}~\cite{subset-error,rust-polonius}.  These errors are
checked at function returns to ensure that loans passed in by the
caller flow back consistently with the region relations declared in
the function signature.  We discuss these checks in the next section.

 \section{A Forward Abstract Interpreter for Polonius Borrow
  Checking}\label{sec:abstract-semantics}

\subsection{Overview of the Abstract Interpreter}

\begin{figure}[htbp]
  \small
  \begin{mathpar}   
  \inferrule[Pol-Fwd-Invalid]{
    p^{\omega} \textbf{ invalidated at } n\quad
    p^{\omega} \in \astore[r] \textbf{ at } n
  }{
    \astore[r] = \invalidls \textbf{ at } \succ(n)
  }

  \inferrule[Pol-Fwd-Check]{
    p \textbf{ accessed at } n\quad
    r \in \typeof{p}\quad
    \astore[r] = \invalidls \textbf{ at } n
  }{
    \textbf{borrow-error at } n
  }
\end{mathpar}
\caption{Inference rules of region invalidation and error detection for forward-only borrow checking}
\label{fig:forward-rules}

\end{figure}

Although rule-based Polonius is more suitable for formalization than the original algorithm, it is still too complicated because it requires mixing forward propagation of loans and backward analysis of liveness.
We observe that region liveness is used only to anticipate future uses
and report errors \emph{eagerly}. It does not affect loan analysis.
Therefore, we could turn rule-based Polonius into forward-only analysis by reporting errors \emph{lazily}. More specifically, when a loan is invalidated, we mark the regions containing it as invalid (called \emph{region invalidation}) as shown in \Rule{Pol-Fwd-Invalid} in~\figref{fig:forward-rules}. An error is reported only when a later operation accesses a place whose type contains an invalid region, as shown in \Rule{Pol-Fwd-Check}.
In summary, the rules in~\figref{fig:forward-rules} replace \Rule{Pol-Loan-Live} and \Rule{Pol-Borrow-Check}. Borrow errors are now reported when accessing invalidated regions.

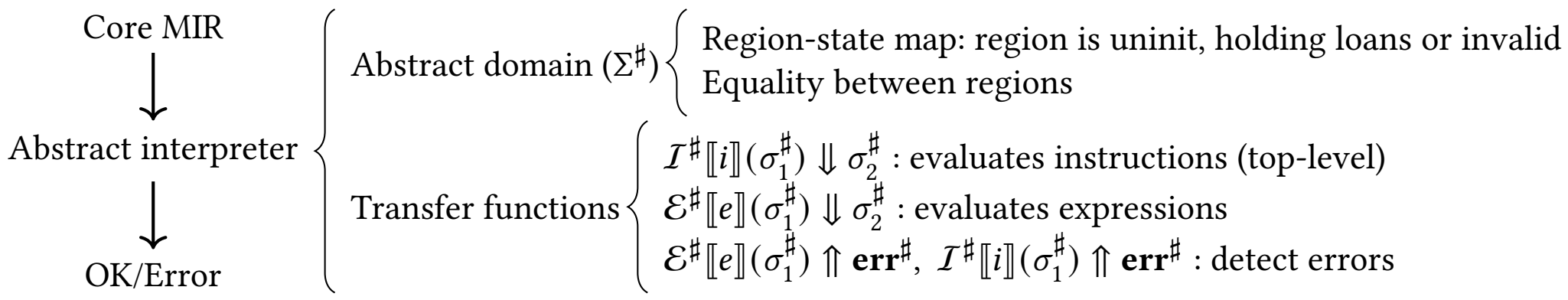
\begin{figure}[t]
	\centering
	\resizebox{\linewidth}{!}{
		\begin{tikzpicture}[
			flow/.style={align=center, font=\small},
			comp/.style={align=left, font=\small},
			item/.style={align=left, font=\small},
			arr/.style={->, thick},
			brace/.style={decorate, decoration={brace, mirror, amplitude=5pt}},
			bracepos/.style={inner sep=0pt, outer sep=0pt},
			]
			
			\node[bracepos] (main-brace) {};
			\node[comp, right=0cm of main-brace] (domain)
			{Abstract domain ($\Astore$)};
			\node[bracepos, right=0.1cm of domain] (domain-brace) {};
			\node[item, right=0cm of domain-brace] (domain-items)
			{\begin{tabular}{@{}l@{}}
					Region-state map: region is uninit, holding loans or invalid\\
					Equality between regions
			\end{tabular}};
			
			\node[bracepos, below=0.75cm of domain-items.south west]
			(operator-row) {};
			\node[comp, anchor=west] (operator)
			at (domain.west |- operator-row) {Transfer functions};
			\node[bracepos, right=0.1cm of operator] (operator-brace) {};
			\node[item, right=0cm of operator-brace] (operator-items)
			{\begin{tabular}{@{}l@{}}
$\Ievals{i}{\astore_1}\Downarrow\astore_2:$ evaluates instructions (top-level)\\
					$\Eevals{e}{\astore_1}\Downarrow\astore_2:$ evaluates expressions\\          
					$\Eevals{e}{\astore_1}\Uparrow\aerr,\app \Ievals{i}{\astore_1}\Uparrow\aerr:$ detect errors\end{tabular}};
			
			\node[fit=(domain)(domain-items),
			inner xsep=0.08cm, inner ysep=0.06cm]
			(domain-block) {};
			\node[fit=(operator)(operator-items),
			inner xsep=0.08cm, inner ysep=0.06cm]
			(operator-block) {};
			
			\node[fit=(domain-items), inner xsep=0.08cm, inner ysep=0.06cm]
			(domain-group) {};
			\node[fit=(operator-items),
			inner xsep=0.08cm, inner ysep=0.06cm]
			(operator-group) {};
			
\draw[brace] ([yshift=-3pt]main-brace.center |- domain-group.north)
			-- ([yshift=3pt]main-brace.center |- operator-group.south) node [midway, left=0.2cm] (interp) {\small Abstract interpreter};
			
			\draw[brace] ([yshift=-3pt]domain-brace.center |- domain-group.north)
			-- ([yshift=3pt]domain-brace.center |- domain-group.south);
			\draw[brace] ([yshift=-3pt]operator-brace.center |- operator-group.north)
			-- ([yshift=3pt]operator-brace.center |- operator-group.south);
			
\node[flow, above=0.6cm of interp] (mir) {Core MIR};
			\node[flow, below=0.6cm of interp] (res) {OK/Error};
			
\draw[arr] (mir) -- (interp);
			\draw[arr] (interp) -- (res);
		\end{tikzpicture}
	}
	\caption{Architecture of the abstract interpreter for Polonius}
	\label{fig:abstract-interpreter-structure}
\end{figure}

In the new rules, we use an abstract state $\astore$ to mark validity of regions and to store its set of loans. It indicates that the forward-only Polonius could be formalized as forward abstract interpretation on an abstract domain. This makes its verification manageable by adopting abstract interpretation techniques. It also gives a runnable algorithm using fixed-point solvers.
The architecture of this abstract interpreter is shown
in~\figref{fig:abstract-interpreter-structure}.
It computes an abstract state $\astore\in\Astore$ at each execution point.
$\astore$ maps each region $r$ to a \emph{region state} $\astore[r]$.  A region state can be uninitialized, valid with a loan set (same as that obtained by constraint solving in Polonius), or invalid.
The transfer functions model transitions between abstract states and detect borrow errors.

The main challenge in designing this abstract interpreter is to faithfully implement the rules of Polonius in the transfer functions and abstract domain.
As shown in~\secref{ssec:polonius-constraint}, Polonius can be viewed as a
constraint-based pointer analysis. Regions are abstract pointers
storing loans that represent borrowed memory locations. This suggests
the naive definition of an abstract state as a mapping from regions to loan sets, modeling constraints like $\rawset{p^\omega}\subseteq r$ as $p^{\omega} \in \astore[r]$.
For example, we replace $\rawset{p^\omega}\subseteq r$ in \Rule{Pol-Loan-Live} with $p^{\omega} \in \astore[r]$ in \Rule{Pol-Fwd-Invalid}.
However, besides such \emph{consumption} of subset constraints between regions, Polonius also \emph{generates} new subset constraints, some of which even represent equality between regions and need to be maintained across program points. How should these constraints be translated into abstract interpretation?

Our key observation is that these subset constraints could be classified into \emph{transient} and \emph{persistent} constraints. The former is modeled as computation of data-flow in transfer functions, while the latter as a more sophisticated abstract domain that partitions regions into equivalence classes.

\myparagraph{Transient constraints}  These are constraints
$r_1\subseteq r_2$ where at least one of $r_1$ and $r_2$ is a region
of a borrow expression (i.e., a temporary region).  Examples include the constraints involving
the borrow expression's temporary region $r$ in \Rule{Pol-Bor-Expr} (e.g., $\rawset{p_1^{\omega}} \subseteq r$) and
\Rule{Pol-Assign}.
We call them transient because here $r_1\subseteq r_2$ only serves to propagate loans in $r_1$ into $r_2$ at a single program point, without requiring the subset relations to be maintained in subsequent execution.  Thus, they can be modeled directly as dataflow
operations in the transfer functions.

\myparagraph{Persistent constraints} These are constraints
$r_1\subseteq r_2$ where both regions appear in the types of
places (i.e., not temporary regions). They arise from Rust's subtyping rules~\cite{rust-subtyping} and are important for operations such as indirect updates
through nested references. 
Examples include region equality introduced by \Rule{Pol-Sub-Mut}
(invariance) and subset relations introduced by \Rule{Pol-Sub-Shr}
(covariance).\footnote{Rust also supports contravariance.
Since our language excludes higher-order functions, we do not
consider contravariance.}
Because these constraints do not involve temporary regions, they need to be maintained during execution, hence should be persistently recorded in the abstract domain.
A naive approach is to use a
relational abstract domain that records both subset and equality relations.
However, we observe that subset relations introduced by shared
references need not be maintained persistently. Since values cannot be updated through shared references during execution (i.e., cannot perform
indirect updates), they can be treated as propagation of loans like transient constraints.
Consequently, we only need to maintain region equality introduced by \Rule{Pol-Sub-Mut} in the abstract domain, which we model by dividing the domain of $\astore$ into equivalence classes of regions modulo this equality.

\subsection{Formalization of the Abstract Interpreter}

\begin{algorithm}[!htbp]
\caption{Borrow Checking}
  \label{alg:borrow-check}
  \setlength{\columnsep}{0.8em}
  \algrenewcommand\algorithmicindent{0.8em}
  \begin{multicols}{2}
  \begin{algorithmic}[1]
    \Procedure{BorrowCheck}{$\prog$}
      \ForAll{$f\in\mathsf{functions}(\prog)$}
        \State $A_f\gets\text{\textsc{Interpret}}(f)$
        \ForAll{$(n,i)\in\mathsf{nodes}(g_f)$}
          \If{$\exists\aerr,\app\Ievals{i}{A_f(n)}\Uparrow\aerr$}
            \State \textbf{return} $\mathsf{Error}$
          \EndIf
        \EndFor
      \EndFor
      \State $\mathcal{A}\gets
        \{f\mapsto A_f\mid f\in\mathsf{functions}(\prog)\}$
      \State \textbf{return} $\mathsf{OK}(\mathcal{A})$
    \EndProcedure
  \end{algorithmic}
  \columnbreak
  \begin{algorithmic}[1]
    \Procedure{Interpret}{$f$}
      \State $\mathbb{L}_{\Astore}\gets
        (\Astore,\sqsubseteq_{\Astore},\sqcup_{\Astore},\bot_{\Astore})$
        \State $\astore\gets\InitFunction{f}$ \Comment{See \figref{fig:abstract-call-and-init}}
        \State $A_f \gets \solver(g_f,\mathbb{L}_{\Astore},
        \Ievals{\cdot}{\cdot},\astore)$
      \State \textbf{return} $A_f$
    \EndProcedure
    \Procedure{$\mathcal{I}^{\sharp}\sem{i}$}{$\astore$}
      \Match{$i$}
        \State $\ldots$ \Comment{See \figref{fig:abstract-evaluation-rules}}
      \EndMatch
    \EndProcedure
  \end{algorithmic}
  \end{multicols}
\end{algorithm}

\algoref{alg:borrow-check} implements our borrow checker for a program
$\prog$. For each function $f$, $\textsc{Interpret}(f)$ computes a map
$A_f$ that assigns an abstract state to each CFG node.
The checker then evaluates each instruction $i$ at node $n$ again to
detect errors, written as $\Ievals{i}{A_f(n)}\Uparrow\aerr$.  If no
instruction raises an error, it returns $\mathsf{OK}(\mathcal{A})$,
collecting the analysis results. Otherwise, it returns
$\mathsf{Error}$.

The $\textsc{Interpret}$ function invokes a dataflow solver to compute
the abstract states of $f$.  The solver evaluates instructions in the
CFG $g_f$ until a fixpoint is reached. It takes a semi-lattice
$\mathbb{L}_{\Astore}$, a transfer function
$\Ievals{\cdot}{\cdot}: I\to\Astore\to(\Astore+\aerr)$, and an initial
state at function entry.
The solver must return a solution 
satisfying the dataflow constraints: for every CFG edge $n\to s$, if
$\Ievals{i_n}{A_f(n)}\Downarrow\astore_s$, then
$\astore_s\sqsubseteq_{\Astore}A_f(s)$.  Here, $i_n$ is the
instruction at node $n$.  This means that $A_f(s)$ must (over-)approximate
the result of evaluating $i_n$ from $A_f(n)$.
We use the classic Kildall's algorithm as the solver as it has already been formally verified in CompCert.
Since $\Ievals{\cdot}{\cdot}$ may raise borrow errors, we use a
\emph{logging monad} as in~\cite{verified-static-analyzer}, to prevent
errors from interrupting the solver's fixpoint computation.  Errors
are recorded in a log while the analysis continues.
After the analysis, errors can be detected by evaluating the
instructions again and inspecting the log.

\begin{figure}[t]
  \small
  \arraycolsep=1pt
  \centering
  \ruleline{\sffamily\textbf{Abstract States}} \[
  \begin{array}[t]{lrll}
    & \ln\in\Ln ::=\app && \text{loan} \\
    & & p^{\omega} & \text{internal loan} \\
    & & \eloan{\rho} & \text{external loan} \\
    & \ls\in & \mathcal{P}(\Ln) & \text{set of loans} \\
    & \rs\in\Rs ::=\app && \text{region state} \\
    & & \bot_\Rs & \text{uninit state} \\
    & & \valid{\ls} & \text{valid state} \\
    & & \invalidls & \text{invalid state}
  \end{array}
  \app
  \begin{array}[t]{lrll}
    & u\in\app & U & \text{union-find} \\
    & \astore,(u,m)\in\Astore ::=\app & U\times(R\to\Rs) & \text{abstract states} \\
    & \aerr\in \aErr ::=\app && \text{reported error} \\
    & & \berrs & \text{borrow error} \\
    & & \subseterrs & \text{illegal region relations}\\[0.5em]
    & \mathbb{L}_{\Rs}::= & (\Rs, \sqsubseteq_\Rs, \sqcup_\Rs, \bot_\Rs, \top_\Rs) & \text{lattice of $\Rs$}\\
    & \mathbb{L}_{\Astore}::= & (\Astore,\sqsubseteq_{\Astore},\sqcup_{\Astore},\bot_{\Astore}) & \text{semi-lattice of \Astore}
  \end{array}  
\]

\[
    \top_\Rs = \invalidls\quad
  \bot_\Rs \sqsubseteq \valid{\ls} \sqsubseteq \invalidls \quad
  \valid{\ls_1} \sqsubseteq \valid{\ls_2} \iff \ls_1 \subseteq \ls_2\]
\[
  \bot \sqcup \rs = \rs \sqcup \bot = \rs\quad
  \invalidls \sqcup \rs = \rs \sqcup \invalidls =\invalidls\quad
  \valid{\ls_1}\sqcup \valid{\ls_2}=\valid{\ls_1\cup \ls_2}
\]
\ruleline{\sffamily\textbf{Operations}} \[
  \mathsf{find}_U:U\times R\to R
  \quad [r]_u:=\mathsf{find}_U(u,r) 
  \quad \mathsf{union}_U:U\times R\times R\to U
  \quad r_1\equiv_u r_2\iff[r_1]_u=[r_2]_u
\]
\[
  \pmb{\lmget{(u,m)}{r}}:=
    \begin{cases}
      m[[r]_u] & \text{if }[r]_u\in\mathsf{dom}(m)\\
      \bot_\Rs & \text{otherwise}
    \end{cases}
  \qquad
  \pmb{\lmset{(u,m)}{r}{\rs}}:=(u,m[[r]_u\leftarrow\rs])
\]
\[
\pmb{\lmunion{(u,m)}{r_1}{r_2}}
    := (u',m[[r_1]_{u'}\leftarrow
      \lmget{(u,m)}{r_1}\sqcup_\Rs\lmget{(u,m)}{r_2}]),
    \quad \text{where }u'=\mathsf{union}_U(u,r_1,r_2)
\]  
  \caption{Definitions of the abstract domain and its operations.  The
    bottom part of \textbf{Abstract State} defines the lattice
    operators for region states (omit $\Rs$ subscript). The components
    of $\mathbb{L}_{\Astore}$ can be simply obtained by lifting those
    in the lattice structures of the union-find and $\Rs$. In
    \textbf{Operations}, the first line shows the standard union-find
    operations. The \textbf{bold} operators define lookup, update, and
    union on the abstract state. We also write operations on $(u,m)$
    directly on $\astore$. For example, $\lmget{\astore}{r}$
    abbreviates $\lmget{(u,m)}{r}$ when $\astore=(u,m)$. }
  \label{fig:abs-dom}
\end{figure}

\myparagraph{Abstract domain}
\figref{fig:abs-dom} defines the abstract state and its operations.
We introduce \emph{external loans} $\eloan{\rho}$ to represent loans
passed from the caller to the callee. Since the callee is analyzed
without knowing the caller's specific loans, these loans are
summarized by a region parameter $\rho$. They are called ``placeholder
loans'' in the Polonius
implementation~\cite{rust-polonius,subset-error}.
External loans are essential for supporting modular analysis.  We
explain their role in function calls later.
An abstract state $\astore$ is a pair $(u,m)$, where $u$ is a
union-find structure representing region equality, and $m$ maps
regions to region states.
The map $m$ stores only one state for each equivalence class in $u$.
We write $[r]_u$ for its representative, computed by
$\mathsf{find}_U(u,r)$.  Looking up or updating a region $r$ accesses
the state of $[r]_u$.  The union operation merges two equivalence
classes and joins their region states.

\begin{figure}[!htbp]
  \smaller
  \captionsetup{skip=7pt}
  \[p_1\#p_2 := \exists q,\app p_1 = q[p_2] \vee p_2 = q[p_1]\]
  \[
    \AllRegs{p} := \pset{r\in R}{p = q[x] \wedge r\in \typeof{x}} \quad
    \RefRegs{p}
      :=\pset{r'}{\exists p',\app
        (p':\&\,r'\,\_\,\_)\wedge\prefix(\deref p',p)}\]
  \[
     \InvalidateLoan{p_a}{\akwrite}{\astore_1} = \astore_2 \quad \text{s.t. } \forall r\app\ls\app (p_b^\omega \in \ls),\app
        \lmget{\astore_1}{r}=\valid{\ls} \imply
        p_a\#p_b\imply
        \lmget{\astore_2}{r}=\invalidls
      \]
  \[
     \InvalidateLoan{p_a}{\akread}{\astore_1} = \astore_2 \quad \text{s.t. } \forall r\app\ls\app (p_b^\mutx \in \ls),\app
        \lmget{\astore_1}{r}=\valid{\ls} \imply
        p_a\#p_b\imply
        \lmget{\astore_2}{r}=\invalidls
  \]
  \[
    \KillLoans{\astore_1}{p}=\astore_2 \quad\text{s.t. }\forall
    r\app\ls,\app \lmget{\astore_1}{r}=\valid{\ls}\imply
    \lmget{\astore_2}{r}=\valid{ \ls\setminus\pset{q[p]^\omega}{\deref^r\in q}}
  \]

  \begin{mathpar}
    \inferrule[P-Abs-OK]{
      \forall r\in\RefRegs{p},\app
      \lmget{\astore}{r}\neq\invalidls \quad
      \bigsqcup\nolimits_{r\in\RefRegs{p}}\lmget{\astore}{r} = \valid{\ls}
}{
      \Pevals{p}{\omega}{\astore}\Downarrow (\rawset{p^\omega} \cup \ls)
    }

    \inferrule[P-Abs-Invalid]{
      \exists r\in\RefRegs{p},\app
        \lmget{\astore}{r}=\invalidls
    }{
      \Pevals{p}{\omega}{\astore}\Uparrow\berrs
    }
  \end{mathpar}
  \caption{Auxiliary functions, region invalidation, loan killing and evaluation of
    places}
  \label{fig:abstract-evaluation-helpers}
  \label{fig:abstract-place-evaluation}

  \captionsetup{type=figure}
  \begin{mathpar}
    \inferrule[E-Abs-Move]{
      \InvalidateLoan{p}{\akwrite}{\astore_1} = \astore_2\\\\
      \forall r\in\AllRegs{p},\app
        \lmget{\astore_2}{r}\neq\invalidls
    }{
      \Eevals{\move{p}}{\astore_1} \Downarrow \astore_2
    }

    \inferrule[E-Abs-Borrow]{
      \mathsf{access} = \mathbf{if}\ \omega=\mathtt{shr}\
      \mathbf{then}\ \akread\ \mathbf{else}\ \akwrite\\\\
      \InvalidateLoan{p}{\mathsf{access}}{\astore_1} = \astore_2 \\
      \forall r'\in\AllRegs{p},\app
        \lmget{\astore_2}{r'}\neq\invalidls \\\\
      \Pevals{p}{\omega}{\astore_2} \Downarrow \ls \\
      \lmset{\astore_2}{r}{\valid{\ls}} = \astore_3
    }{
      \Eevals{\tref[r][\omega]{p}}{\astore_1} \Downarrow \astore_3
    }
  \end{mathpar}
  \caption{Evaluation of move and borrow expressions; operand
    expressions are expanded directly.}
  \label{fig:abstract-expression-evaluation}

\captionsetup{type=figure}
  \begin{mathpar}
    \inferrule[Flow-Loan-Ref]{
      \FlowLoan{\astore_1}{r_s}{r_t}{\nu}
        =\astore_2 \\\\
      \nu'=\mathbf{if}\ \omega=\mathtt{shr}\ \mathbf{then}\ \nu\ \mathbf{else}\ \Invariant
      \\\\
      \FlowLoans{\astore_2}{\tau_s}{\tau_t}{\nu'}
        =\astore_3
    }{
      \FlowLoans{\astore_1}
        {\tref[r_s][\omega]{\tau_s}}
        {\tref[r_t][\omega]{\tau_t}}{\nu}
        =\astore_3
    }

    \inferrule[Flow-Cov]{
      \lmset{\astore_1}{r_t}
        {\lmget{\astore_1}{r_s}
          \sqcup_\Rs\lmget{\astore_1}{r_t}}=\astore_2
    }{
      \FlowLoan{\astore_1}{r_s}{r_t}{\Covariant}
        =\astore_2
    }

    \inferrule[Flow-Inv]{
      \lmunion{\astore_1}{r_s}{r_t}=\astore_2
    }{
      \FlowLoan{\astore_1}{r_s}{r_t}{\Invariant}
        =\astore_2
    }
  \end{mathpar}
  \caption{Operations of flowing loans between reference types, where
    $\nu\in\{\Covariant \text{ (covariant)},\Invariant \text{
      (invariant)}\}$}
  \label{fig:abstract-loan-flow}

  \captionsetup{type=figure}
  \begin{mathpar}
    \inferrule[I-Abs-Assign]{
      \Eevals{e}{\astore_1} \Downarrow \astore_2 \\
      \InvalidateLoan{p}{\akwrite}{\astore_2}
        = \astore_3 \\\\
      \Pevals{p}{\texttt{mut}}{\astore_3} \Downarrow \_ \\
      \KillLoans{\astore_3}{p} = \astore_4 \\\\
\FlowLoans{\astore_4}{\typeof{e}}{\typeof{p}}{\Covariant}
        = \astore_5
    }{
      \Ievals{(\sassign{p}{e},\_)}{\astore_1} \Downarrow \astore_5
    }
\quad
    \inferrule[I-Abs-Call]{
      \Eevalsvec{\op}{\astore_1} \Downarrow \astore_2 \quad
      \TransferCall{\astore_2}{\typeofvec{\op}}{\typeof{f'}}
        \Downarrow \astore_3 \\\\
      \InvalidateLoan{p}{\akwrite}{\astore_3} = \astore_4 \quad
      \Pevals{p}{\texttt{mut}}{\astore_4} \Downarrow \_ \\\\
      \KillLoans{\astore_4}{p} = \astore_5 \quad
      \FlowLoans{\astore_5}{\retty{f'}}{\typeof{p}}{\Covariant}
        = \astore_6
    }{
      \Ievals{(p:=id_{f'}(\overrightarrow{\op}),\_)}{\astore_1}
        \Downarrow \astore_6
    }

    \inferrule[I-Abs-Drop]{
      \InvalidateLoan{p}{\akwrite}{\astore_1} = \astore_2\\\\
      \Pevals{p}{\texttt{mut}}{\astore_2} \Downarrow \_ \\
    }{
      \Ievals{\drop{p}}{\astore_1} \Downarrow \astore_2
    }
	\quad
    \inferrule[I-Abs-Return]{
      \overrightarrow p=
        f.\mathsf{vars}\mathbin{++}[\retx]\mathbin{++}f.\mathsf{params} \\\\
      \Eevals{\move{\retx}}{\astore_1}\Downarrow\astore_2 \\\\
\InvalidateLoanvec{p}{\akwrite}{\astore_2}
      =\astore_3 \\\\      
\CheckReturn{f}{\astore_3}\Downarrow\mathsf{OK}
    }{
      \Ievals{\sreturn{\retx}}{\astore_1}\Downarrow\astore_3
    }
	\quad
    \inferrule[Check-Ret-Region-Err]{
      f=\texttt{fn}\app\id_f
        \langle\overrightarrow{\rho},
          \overrightarrow{\rsubset{\rho_1}{\rho_2}}\rangle\mydots
      \\\\
      \forall\rho\in\overrightarrow{\rho},\app
        \lmget{\astore}{\rho}\neq\invalidls
      \\\\
      \rho_s,\rho_t\in\overrightarrow{\rho} \quad
      \rho_s\neq\rho_t \quad
      \lmget{\astore}{\rho_t}=\valid{\ls}
      \\\\
      \eloan{\rho_s}\in\ls \quad
      (\rho_s,\rho_t)\notin\overrightarrow{\rsubset{\rho_1}{\rho_2}}
    }{
      \CheckReturn{f}{\astore}\Uparrow\subseterrs
    }
  \end{mathpar}
  \caption{Evaluation of assignments, drops, function calls and
    function returns ($f$ is the current function)}
  \label{fig:abstract-evaluation-rules}

  \captionsetup{type=figure}
  \begin{mathpar}
    \inferrule[Transfer-Call-OK]{
      \tau_f=\texttt{fn}\app
        \langle\overrightarrow{\rho},
          \overrightarrow{\rsubset{\rho_1}{\rho_2}}\rangle
        (\overrightarrow{\tau_{\mathsf{par}}})\rightarrow\tau
      \quad
      \FlowLoansvec{\astore_1}{\tau_{\mathsf{arg}}}{\tau_{\mathsf{par}}}{\Covariant}=\astore_2
      \\\\
      \forall\rho\in\overrightarrow{\rho},\app
        \lmget{\astore_2}{\rho}\neq\invalidls
      \\
      \overrightarrow{\mathsf{Flow}}
        (\astore_2,\overrightarrow{\rho_1},\overrightarrow{\rho_2},\Covariant)
        =\astore_3
    }{
      \TransferCall{\astore_1}{\overrightarrow{\tau_{\mathsf{arg}}}}{\tau_f}
        \Downarrow\astore_3
   }
      
    \inferrule[Init-Function]{
      f = \texttt{fn}\app \id_f
        \langle \overrightarrow{\rho},
          \overrightarrow{\rsubset{\rho_1}{\rho_2}}\rangle \mydots
      \quad
      \astore_0=\bot_{\Astore}\\\\
      \lmsetvec{\astore_0}{\rho}{\valid{\rawset{\eloan{\rho}}}}
        =\astore_1
    }{
      \InitFunction{f}=\astore_1
    }
  \end{mathpar}
  \caption{Operations of analyzing a function call and function
    initialization}
  \label{fig:abstract-call-and-init}
\end{figure}

\Needspace{16\baselineskip}
\begingroup
\setlength{\intextsep}{0pt}
\begin{wraptable}{r}{0.35\textwidth}
  \raggedleft
  \caption{Design of transfer functions}
  \label{tab:transfer-polonius-rules}
  \smaller
  \renewcommand{\Rule}[1]{\hyperlink{#1}{\textsc{#1}}}
  \setlength{\tabcolsep}{3pt}
  \begin{tabular}{@{}l l@{}}
    \toprule
    \textbf{Transfer functions} & \textbf{Polonius rules} \\
    \midrule
    \Rule{I-Abs-Assign} & \Rule{Pol-Assign} \\
    \midrule
    \begin{tabular}{@{}l@{}}
      \Rule{E-Abs-Borrow}\\
      \Rule{P-Abs-OK}
    \end{tabular} & \Rule{Pol-Bor-Expr} \\
    \midrule
    \Rule{Flow-Loan-Ref} & \\
    \quad \Rule{Flow-Cov} & \hyperlink{Pol-Sub-Shr}{\textsc{Pol-Sub-Mut/Shr}} \\
    \quad \Rule{Flow-Inv} & \Rule{Pol-Sub-Mut} \\
    \midrule
    \hyperref[fig:abstract-evaluation-helpers]{$\mathsf{Invalidate}^{\sharp}$} & \Rule{Pol-Fwd-Invalid} \\
    \midrule
    \hyperref[fig:abstract-evaluation-helpers]{$\mathsf{Kill}^{\sharp}$} & \hyperlink{Pol-Loan-Flow}{$p^\omega\app\textbf{not killed at}\app n$} \\
    \midrule
    \Rule{P-Abs-Invalid} & \Rule{Pol-Fwd-Check} \\
    \bottomrule
  \end{tabular}
\end{wraptable}
\myparagraph{Transfer functions}
We present the main cases of the transfer functions
in~\figref{fig:abstract-evaluation-rules}, including assignments,
drops, function calls, and returns.
\tabref{tab:transfer-polonius-rules} shows how these functions
correspond to the Polonius rules.
Transfer functions model transient constraints by updating the region
states and persistent constraints by updating the union find.
We illustrate them using the examples
from~\secref{ssec:polonius-constraint} and a new function call
example. In these examples, reassignments to \mcode{Box} variables are
expanded into a drop followed by an assignment, as in the actual Core
MIR program. This makes clear which transfer functions are applied.
The comments record the abstract state after each line.
We write
$[r_1,\mydots,r_n]\mapsto
\rawset{p_1^{\omega_1},\mydots,p_k^{\omega_k}}$ to mean that
$r_1,\mydots,r_n$ belong to the same equivalence class and share a
valid region state containing the loan set
$\rawset{p_1^{\omega_1},\mydots,p_k^{\omega_k}}$.
\par
\WFclear
\endgroup
\begin{minted}[xleftmargin=1.5em,formatcom=\RustCodeLineNumbers,mathescape=true,escapeinside=||]{rust}
let mut v = Box::new(1); // ['x]$\mapsto\bot_\Rs$, ['y]$\mapsto\bot_\Rs$, ['r1]$\mapsto\bot_\Rs$, ['r2]$\mapsto\bot_\Rs$  ($\textup{\Rule{I-Abs-Assign}}$)
let mut x: &'x mut Box<i32> = &'r1 mut v;//['x]$\mapsto\rawset{\mcode{v}^{\mutx}}$,['y]$\mapsto\bot_\Rs$,['r1]$\mapsto\rawset{\mcode{v}^{\mutx}}$,$\mydots$ ($\textup{\Rule{I-Abs-Assign}}$)
let mut y: &'y i32 = &'r2 **x; //['x]$\mapsto\rawset{\mcode{v}^{\mutx}}$,['y]$\mapsto\rawset{\mcode{v}^{\mutx},\mcode{\deref\deref x}^{\texttt{s}}}$,['r2]$\mapsto\rawset{\mcode{v}^{\mutx},\mcode{\deref\deref x}^{\texttt{s}}}$,$\mydots$ ($\textup{\Rule{I-Abs-Assign}}$)
drop(v); // ['x]$\mapsto\invalidls$, ['y]$\mapsto\invalidls$, $\mydots$ ($\textup{\Rule{I-Abs-Drop}}$); 
v = Box::new(2);// ['x]$\mapsto\invalidls$, ['y]$\mapsto\invalidls$, $\mydots$ ($\textup{\Rule{I-Abs-Assign}}$)
println!("{}", *y); // borrow-error: the accessed region 'y is invalid ($\textup{\Rule{P-Abs-Invalid}}$)
\end{minted}
\label{exm:borrow-reborrow-abstract}
The results of this example agree with those of the rule-based
Polonius in~\secref{ssec:polonius-constraint}.
Rule \Rule{I-Abs-Assign} models \Rule{Pol-Assign}, while
\Rule{E-Abs-Borrow} models \Rule{Pol-Bor-Expr}.  The latter invokes
$\Pevals{p}{\omega}{\astore}$ to collect the directly issued loan and
loans from the prefix regions.
For example, at line~3,
$\Pevals{\mcode{\deref\deref x}}{\texttt{shr}}{\astore_{l2}}$ computes
$\rawset{\mcode{v}^{\mutx}, \mcode{\deref\deref x}^{\texttt{s}}}$.
Finally, $\mathsf{FlowLoans}$ propagates loans from the regions of the
expression type to those of the assignee type, modeling
\Rule{Pol-Sub-Mut} and \Rule{Pol-Sub-Shr}. At line~3, \Rule{Flow-Cov}
propagates the loan set
$\rawset{\mcode{v}^{\mutx}, \mcode{\deref\deref x}^{\texttt{s}}}$ from
\ticktt{r2} to \ticktt{y} through a join operation.

At line~4, $\mathsf{Invalidate}^\sharp$ in \Rule{I-Abs-Drop}
invalidates \ticktt{x} and \ticktt{y}, which contain loans conflicting
with \mcode{v}.  Errors are detected when a place $p$ is later
accessed, through validity checks over $\RefRegs{p}$ and
$\AllRegs{p}$.
Intuitively, $\RefRegs{p}$ contains the regions of reference prefixes
used to reach $p$, while $\AllRegs{p}$ also includes the regions in
the type of the value stored at $p$.  The former checks whether
locating $p$ encounters an invalid region, as in \Rule{I-Abs-Assign}
and \Rule{I-Abs-Drop}.  The latter also checks the validity of the
value stored at $p$ and the references it contains.
A failed validity check raises $\berrs$, as in \Rule{P-Abs-Invalid}.
We omit the error cases of $\Eevals{\cdot}{\cdot}$ and
$\Ievals{\cdot}{\cdot}$ for brevity.

We use the example of nested references to illustrate how region equality
is established:
\begingroup
\begin{minted}[xleftmargin=1.5em,formatcom=\RustCodeLineNumbers,mathescape=true,escapeinside=||, firstnumber=1]{rust}
// For brevity, we omit the region states for temporary regions, e.g., 'r1 and 'r2.
let mut v1 = 1; let mut v2 = Box::new(2); // ['x]$\mapsto\bot_\Rs$, ['y1]$\mapsto\bot_\Rs$, ['y2]$\mapsto\bot_\Rs$ ($\textup{\Rule{I-Abs-Assign}}$)
let mut x: &'x mut i32 = &'r1 mut v1; // ['x]$\mapsto\rawset{\mcode{v1}^{\mutx}}$, ['y1]$\mapsto\bot_\Rs$, ['y2]$\mapsto\bot_\Rs$ ($\textup{\Rule{I-Abs-Assign}}$)
let mut y: &'y1 mut &'y2 mut i32 = &'r2 mut x; // ['y1]$\mapsto\rawset{\mcode{x}^{\mutx}}$, ['y2,'x]$\mapsto\rawset{\mcode{v1}^{\mutx}}$ ($\textup{\Rule{I-Abs-Assign}}$)
*y = &'r3 mut *v2; // ['y1]$\mapsto\rawset{\mcode{x}^{\mutx}}$, ['y2,'x]$\mapsto\rawset{\mcode{v1}^{\mutx}, \mcode{\deref v2}^{\mutx}}$ ($\textup{\Rule{I-Abs-Assign}}$)
drop(v2); // ['y1]$\mapsto\rawset{\mcode{x}^{\mutx}}$, ['y2,'x]$\mapsto\invalidls$ ($\textup{\Rule{I-Abs-Drop}}$)
v2 = Box::new(3); // ['y1]$\mapsto\rawset{\mcode{x}^{\mutx}}$, ['y2,'x]$\mapsto\invalidls$ ($\textup{\Rule{I-Abs-Assign}}$)
*x = 4; // borrow-error: the accessed region 'x is invalid ($\textup{\Rule{P-Abs-Invalid}}$)
\end{minted}
\label{exm:nested-ref-abstract}
At line~3, $\mathsf{FlowLoans}$ invokes \Rule{Flow-Inv} to merge
\ticktt{x} and \ticktt{y2} into the same equivalence class.  Thus,
updating \ticktt{y2} at line~4 also (indirectly) updates \ticktt{x}.
Note that this update joins the new loan into the existing loan set
rather than overwriting it.  Overwriting occurs only when a variable
is reassigned, similar to killing a variable in liveness analysis. We
leave this detail to the technical report.

We use the following function call example to illustrate how to
support modular analysis:
\begin{minted}[xleftmargin=1.5em,formatcom=\RustCodeLineNumbers,mathescape=true,escapeinside=||]{rust}
// fn choose<'a,'b,'c>(a: &'a mut i32, b: &'b mut i32, d: bool) -> &'c mut i32
//  where 'b:'c { if d { return a; } else { return b; } } // If checker accepts this function
let mut x = Box::new(1); let mut y = Box::new(2); // ['z]$\mapsto\bot_\Rs$ ($\textup{\Rule{I-Abs-Assign}}$)
let z: &'z mut i32 = choose(&'x mut *x, &'y mut *y, true); // ['z]$\mapsto\rawset{\mcode{\deref y}^{\mutx}}$ ($\textup{\Rule{I-Abs-Call}}$)
drop(x); // ['z]$\mapsto\rawset{\mcode{\deref y}^{\mutx}}$ ($\textup{\Rule{I-Abs-Drop}}$); z now points to freed memory, but it is not invalidated.
x = Box::new(2); // ['z]$\mapsto\rawset{\mcode{\deref y}^{\mutx}}$ ($\textup{\Rule{I-Abs-Assign}}$)
println!("{}", *z); // No borrow-error: 'z is valid, but this read accesses freed memory.
\end{minted}
\label{exm:fun-call1}
At line 2, the \mcode{choose} function (its definition shown in the
comments) may return a reference to either \mcode{\deref x} or
\mcode{\deref y}.  For this call, \Rule{I-Abs-Call} evaluates the
arguments, and \Rule{Transfer-Call-OK} propagates their loans to the
parameter regions. It then applies the region relations in the
signature of \mcode{choose} to \emph{simulate} how the loans return from 
callee.
Since the signature contains only \ticktt{b}:\ticktt{c}, only
$\mcode{\deref y}^{\mutx}$ flows to \ticktt{c} and then to \ticktt{z}.
However, the concrete execution returns a reference to \mcode{\deref
  x}. Dropping \mcode{x} therefore frees the memory referenced by
\mcode{z} without invalidating \ticktt{z}.  The subsequent access
through \mcode{z} would access freed memory without raising a borrow
error.
To prevent this, the borrow checker must reject the incomplete
signature of \mcode{choose}, which lacks the region relation
\ticktt{a}:\ticktt{c}.  This relation tells the caller that the loan
$\mcode{\deref x}^{\mutx}$ may flow into \ticktt{z}.  With this
relation, \ticktt{z} would contain the loan after the call.  Dropping
\mcode{x} would then invalidate \ticktt{z}, causing the subsequent
access to raise error.

We now explain how our borrow checker rejects this \mcode{choose}
function by \Rule{Check-Ret-Region-Err}:
\begin{minted}[xleftmargin=1.5em,formatcom=\RustCodeLineNumbers,mathescape=true,escapeinside=||]{rust}
fn choose<'a,'b,'c> ...{// $[\ticktt{a}]\mapsto\loans{\eloan{\ticktt{a}}},\app [\ticktt{b}]\mapsto\loans{\eloan{\ticktt{b}}},\app [\ticktt{c}]\mapsto\loans{\eloan{\ticktt{c}}}$ ($\textup{\Rule{Init-Function}}$)
  if d { return a; } // $[\ticktt{c}]\mapsto\loans{\eloan{\ticktt{c}},\eloan{\ticktt{a}}},\app [\ticktt{a}]\mapsto\loans{\eloan{\ticktt{a}}},\app [\ticktt{b}]\mapsto\loans{\eloan{\ticktt{b}}}$
  else { return b; } // $[\ticktt{c}]\mapsto\loans{\eloan{\ticktt{c}},\eloan{\ticktt{b}}},\app [\ticktt{a}]\mapsto\loans{\eloan{\ticktt{a}}},\app [\ticktt{b}]\mapsto\loans{\eloan{\ticktt{b}}}$
       // Joined state: $[\ticktt{c}]\mapsto\loans{\eloan{\ticktt{c}},\eloan{\ticktt{a}},\eloan{\ticktt{b}}},\app [\ticktt{a}]\mapsto\loans{\eloan{\ticktt{a}}},\app [\ticktt{b}]\mapsto\loans{\eloan{\ticktt{b}}}$
} // By $\textup{\Rule{Check-Ret-Region-Err}}$: 'a:'c and 'b:'c must both exist in the function signature.
\end{minted}
Recall that an external loan $\eloan{\rho}$ summarizes the loans
supplied by the caller through region parameter $\rho$, which are
unknown to the callee.  During the callee's analysis, these external
loans are propagated like ordinary loans.
In the example, the two return branches propagate $\eloan{\ticktt{a}}$
and $\eloan{\ticktt{b}}$ into \ticktt{c}, respectively. Thus, the
joined state of \ticktt{c} contains both external loans.
At function return, $\mathsf{CheckReturn}$ requires $\rho_s:\rho_t$
whenever $\eloan{\rho_s}$ has flowed into $\rho_t$.
\Rule{Check-Ret-Region-Err} reports $\subseterrs$ for this function as
\ticktt{a}:\ticktt{c} is missing. This ensures that the caller can
soundly simulate the callee's loan flow using only the region
relations in the function signature.
\par\endgroup

 \section{Borrow-Aware Semantics}\label{sec:concrete-semantics}

To verify memory safety, we need a concrete semantics that accurately describes when memory errors could happen. Traditional small-step operational semantics suffice for this purpose.
However, it alone is not enough for connecting with the abstract interpreter (i.e., proving its soundness) whose conservative nature means the interpretation could terminate way earlier before the actual memory errors happen because of violation of stacked AXM.
Therefore, the main challenge is to instrument the concrete semantics with a notion of stacked AXM discipline that excludes ill-behaved interleaved accesses through pointers to the same location. The instrumentation does not change the faithfulness of concrete semantics; it only adds the minimum information to enforce~stacked~AXM.

We observe that our forward-only abstract interpreter for Polonius is
already close to concrete program execution.
We therefore design a concrete semantics by concretizing its
abstractions. In particular, regions as abstract pointers storing
loans are concretized as reference values carrying \emph{borrow sets},
which record the precise places borrowed by each reference.  Dynamic
borrow checking uses \emph{reference invalidation} and checks for
invalid references, corresponding to region invalidation and checks
for invalid regions, respectively.
We use a structured, \emph{value-based memory model} based on Rust's
ownership discipline. This allows us to represent borrowed locations
using named places rather than low-level memory addresses, while
supporting reasoning about memory safety.

\subsection{Memory Model}

\begin{figure}[!htbp]
  \smaller
  \arraycolsep=1pt
  \centering
\ruleline{\sffamily\textbf{Concrete States}}
\[
\begin{array}{ll}   
\begin{array}[t]{lrll}

& \ell ::=\app & q[x]\quad(\deref^r\notin q) & \text{move paths}\\
& \bs \in & \mathcal{P}(P) & \text{borrow sets \quad \quad \quad \quad}\\
  
& v \in V           ::=\app && \text{values} \\
&                           & \texttt{true} \stk \texttt{false} \stk \mathit{n_{\kwd{i32}}}\dots & \text{constants} \\
&                           & \eoref{\ell}{\bs} & \text{reference values} \\
&                           & \erefbot & \text{invalid references} \\
&                           & \ebox{v} & \text{heap-allocated values}\\  
&                           & \bot & \text{uninitialized values} \\
&                           & \vstruct{\id}{\rawset{\overrightarrow{\fid \mapsto v}}} & \text{struct values} \\
&                           & \venum{\id}{\fid}{v} & \text{enum values}
\end{array}
\begin{array}[t]{lrll}
  & \cstore \in \Cstore        ::=\app & \emptyset \mid x \mapsto v, \sigma & \text{memory model (store)}\\
  & \kappa \in K ::= \app & \emptyset \mid \kframe{f}{n}{p},\kappa & \text{call stacks} \\
  & \chi \in \Sig ::= \app &&   \text{instruction signals}\\
  &                        & \normalsig{n} & \text{continue at node $n$}\\
  &                        & \callsig{f}{\vec v} & \text{call $f$ with $\vec v$}\\
  &                        & \returnsig{v} & \text{return $v$}\\

& s \in S ::= \app &&\text{program states}\\
&                   &\regstate{f}{n}{\cstore}{\kappa} & \text{regular states}\\
&                   & \borrowerror & \text{borrow-error states}
\end{array}
\end{array}
\]

\ruleline{\sffamily\textbf{Memory Read and Write}}
  \begin{mathpar}
    \inferrule[V-Read]{
      \ell = q[x] \quad
      \cstore[x] = v_x \quad
      \VreadCtx{q}{v_x}{v}
    }{
      \Vread{\ell}{\cstore}{v}
    }

    \inferrule[VR-Base]{ }{
      \VreadCtx{[.]}{v}{v}
    }

    \inferrule[VR-Box]{
      \VreadCtx{q}{v_P}{\ebox{v}}
    }{
      \VreadCtx{(\deref q)}{v_P}{v}
    }

    \inferrule[VR-StructField]{
      \VreadCtx{q}{v_P}
        {\vstruct{\id}{\overrightarrow{\mathit{fv}}}}
    }{
      \VreadCtx{(q.\fid_i)}{v_P}{\overrightarrow{\mathit{fv}}[\fid_i]}
    }

    \inferrule[V-Write]{
      \ell = q[x] \quad
      \cstore[x] = v_x \\\\
      \VwriteCtx{q}{v_x}{v}{v'_x}
}{
      \Vwrite{\ell}{v}{\cstore}{\cstore[x \leftarrow v'_x]}
    }

    \inferrule[VW-Base]{ }{
      \VwriteCtx{[.]}{v_P}{v}{v}
    }

    \inferrule[VW-Box]{
      \VreadCtx{q}{v_P}{\ebox{\_}} \\\\
      \VwriteCtx{q}{v_P}{\ebox{v}}{v'_P}
    }{
      \VwriteCtx{(\deref q)}{v_P}{v}{v'_P}
    }

    \inferrule[VW-StructField]{
      \VreadCtx{q}{v_P}
        {\vstruct{\id}{\overrightarrow{\mathit{fv}}}} \\\\
      \VwriteCtx{q}{v_P}
        {\vstruct{\id}{\overrightarrow{\mathit{fv}}
          [\fid_i\leftarrow v]}}{v'_P}
    }{
      \VwriteCtx{(q.\fid_i)}{v_P}{v}{v'_P}
    }
\end{mathpar}
  \caption{Program states and memory read and write operations.}
  \label{fig:program-state}

\end{figure}

\figref{fig:program-state} presents our structured memory model. Its design is standard for operational semantics except that information about structured memory addresses and borrowing is instrumented into reference values. A reference value $\eoref{\ell}{\bs}$ stores a \emph{move
  path} $\ell$ identifying its target and a borrow set $\bs$ recording the places
it borrows from.  Move paths come from Rust's move
checking~\cite{move-paths} and serve as memory addresses.
This representation relies on the absence of pointer arithmetic and
Rust's ownership discipline.  We can identify locations using
syntactic places, and each owned value has a unique move path.
Updating a value through its move path corresponds to updating the
value at its memory address.
To avoid name conflicts across stack frames, the root variable of each
move path or place is paired with a unique \emph{frame identifier},
which we omit from the notation.  Memory is therefore represented as a
map from uniquely identified variables to structured values.
Program states are used in small-step transitions guided by
signals produced by instruction evaluation.

Read and write operations accept only move paths as their
targets.
We write $\Vread{\ell}{\cstore}{v}$ to mean that reading location
$\ell$ from $\cstore$ produces value $v$.
Similar to traversing a tree, \Rule{V-Read} first reads the value
$v_x$ stored at the root variable $x$ of $\ell$, and then uses the
auxiliary judgment $\VreadCtx{q}{v_x}{v}$ to follow the projection
$q$, viewed as a path in the value tree rooted at $v_x$, and obtain
the value $v$.
For the write operation, we write
$\Vwrite{\ell}{v}{\cstore}{\cstore'}$ to mean that writing value $v$
to location $\ell$ in $\cstore$ produces the updated memory
$\cstore'$.
\Rule{V-Write} first extracts the old value stored at the root
variable of $\ell$.  It then uses the auxiliary judgment
$\VwriteCtx{q}{v_x}{v}{v_x'}$ to follow the projection $q$ and replace
the corresponding subtree of $v_x$ with the new value $v$, producing
an updated root value $v_x'$.  Finally, $v_x'$ is written back to the
root variable in memory.
The rules for enum values are similar and are omitted for
brevity.

\subsection{Operational Semantics}

\begin{figure}[!htbp]
  \smaller

  \begin{mathpar}
    \inferrule[P-Var]{ }{
      \Peval{x}{\cstore} \Downarrow (x,\rawset{x})
    }

    \inferrule[P-Field]{
      \Peval{p}{\cstore} \Downarrow (\ell_1,\bs)
    }{
      \Peval{p.\fid}{\cstore}
        \Downarrow (\ell_1.\fid,\pset{p'.\fid}{p'\in\bs})
    }

    \inferrule[P-Box]{
      \Peval{p}{\cstore} \Downarrow (\ell_1,\bs_1) \quad
      \Vread{\ell_1}{\cstore}{\ebox{\_}}
    }{
      \Peval{\deref p}{\cstore}
        \Downarrow (\deref \ell_1,\pset{\deref p'}{p'\in\bs_1})
    }

    \inferrule[P-MutRef]{
      \Peval{p}{\cstore} \Downarrow (\ell_1,\bs_1) \\\\
      \Vread{\ell_1}{\cstore}{\emref{\ell_2}{\bs_2}}
    }{
      \Peval{\deref p}{\cstore}
        \Downarrow (\ell_2,\pset{\deref p'}{p'\in\bs_1}\cup\bs_2)
    }

    \inferrule[P-ShrRef]{
      \Peval{p}{\cstore} \Downarrow (\ell_1,\bs_1) \\\\
      \Vread{\ell_1}{\cstore}{\esref{\ell_2}{\bs_2}}
    }{
      \Peval{\deref p}{\cstore} \Downarrow (\ell_2,\bs_2)
    }
    
    \inferrule[P-Invalid]{
      \Peval{p}{\cstore} \Downarrow (\ell_1,\bs_1) \\\\
      \Vread{\ell_1}{\cstore}{\erefbot}
    }{
      \Peval{\deref p}{\cstore} \Uparrow \berr
    }

    \inferrule[E-Copy]{
      \InvalidateRef{\ell}{\akread}{\cstore_1} = \cstore_2 \\\\
      \Peval{p}{\cstore_2} \Downarrow (\ell,\_) \quad
      \Vread{\ell}{\cstore_2}{v}
    }{
      \Eeval{\copyp{p}}{\cstore_1} \Downarrow (v,\cstore_2)
    }

    \inferrule[E-Move]{
      \InvalidateRef{\ell}{\akwrite}{\cstore_1} = \cstore_2 \\\\
\Vread{\ell}{\cstore_2}{v} \quad
      \Vwrite{\ell}{\bot}{\cstore_2}{\cstore_3}
    }{
      \Eeval{\move{\ell}}{\cstore_1} \Downarrow (v,\cstore_3)
    }   

    \inferrule[E-Borrow]{
      \mathsf{acc} = \mathbf{if}\ \omega=\mathtt{shr}\
      \mathbf{then}\ \akread\ \mathbf{else}\ \akwrite\\\\
      \InvalidateRef{p}{\mathsf{acc}}{\cstore_1} = \cstore_2 \quad
      \Peval{p}{\cstore_2} \Downarrow (\ell,\bs)
    }{
      \Eeval{\tref[r][\omega]{p}}{\cstore_1}
        \Downarrow (\eoref{\ell}{\bs},\cstore_2)
    }
\end{mathpar}

  \caption{Evaluation of places and expressions}
  \label{fig:place-evaluation}

\[
    \InvalidateRef{p_a}{\akwrite}{\cstore_1}=\cstore_2
    \quad\text{s.t. }\forall\ell\app\bs\app\omega\app(p_b\in\bs),\app
      \Vread{\ell}{\cstore_1}{\eoref{\_}{\bs}}
      \imply p_a\#p_b\imply
      \Vread{\ell}{\cstore_2}{\erefbot}
  \]
  \[
    \InvalidateRef{p_a}{\akread}{\cstore_1}=\cstore_2
    \quad\text{s.t. }\forall\ell\app\bs\app(p_b\in\bs),\app
      \Vread{\ell}{\cstore_1}{\emref{\_}{\bs}}
      \imply p_a\#p_b\imply
      \Vread{\ell}{\cstore_2}{\erefbot}
  \]
  \[
    \killborrefs{\bs_{\mathsf{kill}}}{\cstore_1}=\cstore_2
    \quad\text{s.t. }\forall\ell\app\ell_t\app\bs\app\omega,\app
    \Vread{\ell}{\cstore_1}{\eoref{\ell_t}{\bs}}\imply
    \Vread{\ell}{\cstore_2}{\eoref{\ell_t}{(\bs\setminus\pset{q[p_k]}{
          p_k\in\bs_{\mathsf{kill}}\wedge\deref^r\in q})}}
  \]
  \caption{Reference invalidation and killing borrow sets}
  \label{fig:reference-invalidation}

\begin{mathpar}
    \inferrule[I-Assign]{
      \Eeval{e}{\cstore_1} \Downarrow (v,\cstore_2) \\
      x \fresh \dom(\cstore_2)\\
      \cstore_3 = x\mapsto v,\cstore_2\\\\
      \InvalidateRef{p}{\akwrite}{\cstore_3} = \cstore_4\\
      \Peval{p}{\cstore_4} \Downarrow (\ell, \bs)\\\\
      \killborrefs{\bs}{\cstore_4} = \cstore_5 \\
      \Vwrite{\ell}{\cstore_5[x]}{\cstore_5}{\cstore'_5}\\
      \cstore_6 = \cstore'_5\setminus(x\mapsto\_)
    }{
      \Ieval{(\sassign{p}{e},n)}{\cstore_1}
        \Downarrow (\normalsig{n},\cstore_6)
    }

    \inferrule[I-Drop]{
      \InvalidateRef{p}{\akwrite}{\cstore_1} = \cstore_2 \\\\
      \Peval{p}{\cstore_2} \Downarrow (\ell,\_) \\
      \Vwrite{\ell}{\bot}{\cstore_2}{\cstore_3}
    }{
      \Ieval{(\drop{p},n)}{\cstore_1}
      \Downarrow (\normalsig{n},\cstore_3)
    }         
\end{mathpar}

\begin{mathpar}
\inferrule[Step-Normal]{
      \Ieval{g_f(n)}{\cstore_1} \Downarrow (\normalsig{n'},\cstore_2)
    }{
      \regstate{f}{n}{\cstore_1}{\kappa}
      \rightarrow
      \regstate{f}{n'}{\cstore_2}{\kappa}
    }

    \inferrule[Step-Call]{
      \Ieval{g_f(n)}{\cstore_1} \Downarrow (\callsig{f'}{\vec v},\cstore_2)\quad
      \initframe{f'}{\vec v}{\cstore_2}\Downarrow\cstore_3
    }{
      \regstate{f}{n}{\cstore_1}{\kappa}
      \rightarrow
\regstate{f'}{\entry{g_{f'}}}{\cstore_3}{\kframe{f}{\succ(n)}{p} :: \kappa}
    }

    \inferrule[Step-Return]{
      \Ieval{g_f(n)}{\cstore_1}\Downarrow(\returnsig{v},\cstore_2)\quad
      \popstack{f}{\cstore_2}\Downarrow\cstore_3\\\\
\Ieval{(\sassign{p}{v},n')}{\cstore_3}
        \Downarrow (\normalsig{n'},\cstore_4) \app(\checkgray{\text{assign return value}})
    }{
      \regstate{f}{n}{\cstore_1}
        {\kframe{f'}{n'}{p} :: \kappa'}
      \rightarrow
      \regstate{f'}{n'}{\cstore_4}{\kappa'}
    }
    
    \inferrule[Step-Bor-Err]{
      \Ieval{g_f(n)}{\cstore} \Uparrow \berr
    }{
      \regstate{f}{n}{\cstore}{\kappa}
      \rightarrow
      \borrowerror
    }       
  \end{mathpar}
  \caption{Evaluation of instructions, initialization and removal of
    stack frames, and small-step semantics}
  \label{fig:small-step}
\end{figure}

Each abstract operator in the abstract interpreter has a concrete
counterpart in our borrow-aware semantics.
\figref{fig:place-evaluation} presents place and expression
evaluation, \figref{fig:reference-invalidation} defines reference
invalidation, and \figref{fig:small-step} presents instruction
evaluation and the small-step semantics.
We use the same three examples from the previous section to explain
these rules and compare the concrete results with those of the
abstract interpreter to illustrate the approximations made by the
analysis.
\begin{minted}[xleftmargin=1.5em,formatcom=\RustCodeLineNumbers,mathescape=true,escapeinside=||]{rust}
let mut v = Box::new(1); // $\mcode{v}\mapsto\ebox{1},\app \mcode{x}\mapsto\bot,\app \mcode{y}\mapsto\bot$ ($\textup{\Rule{I-Assign}}$)
let mut x: &'x mut Box<i32> = &'r1 mut v; // $\mcode{v}\mapsto\ebox{1},\app \mcode{x}\mapsto\emref{\mcode{v}}{\rawset{\mcode{v}}},\app \mcode{y}\mapsto\bot$ ($\textup{\Rule{I-Assign}}$)
let mut y: &'y i32 = &'r2 **x; // $\mcode{v}\mapsto\ebox{1},\app \mcode{x}\mapsto\emref{\mcode{v}}{\rawset{\mcode{v}}},\app \mcode{y}\mapsto\esref{\mcode{\deref v}}{\rawset{\mcode{\deref\deref x},\mcode{\deref v}}}$ ($\textup{\Rule{I-Assign}}$)
drop(v); // $\mcode{v}\mapsto\bot,\app \mcode{x}\mapsto\erefbot,\app \mcode{y}\mapsto\erefbot$ ($\textup{\Rule{I-Drop}}$)
v = Box::new(2); // $\mcode{v}\mapsto\ebox{2},\app \mcode{x}\mapsto\erefbot,\app \mcode{y}\mapsto\erefbot$ ($\textup{\Rule{I-Assign}}$)
println!("{}", *y); // borrow-error: y stores an invalid reference ($\textup{\Rule{P-Invalid}}$)
\end{minted}
\label{exm:borrow-reborrow-concrete}
The comments show the concrete memory state after each line.
At line~3, \Rule{I-Assign} evaluates the borrow expression using
\Rule{E-Borrow} and stores the result in a fresh temporary variable
before assigning it to \mcode{y}.
The temporary variable allows
\hyperref[fig:reference-invalidation]{$\mathsf{Invalidate}$} and
\hyperref[fig:reference-invalidation]{$\mathsf{Kill}$} to handle the
evaluated value like other values in memory.
To compute the target move path and borrow set for \mcode{y},
\Rule{E-Borrow} invokes $\Peval{\mcode{\deref\deref x}}{\cstore}$.
This applies \Rule{P-Var}, \Rule{P-MutRef}, and \Rule{P-Box} in
sequence, producing
$\Peval{\mcode{\deref\deref x}}{\cstore}\Downarrow
(\mcode{\deref v},
\rawset{\mcode{\deref\deref x},\mcode{\deref v}})$.
The resulting reference therefore sharedly borrows
\mcode{\deref\deref x} and \mcode{\deref v}.
This computation corresponds to abstract place evaluation
$\Pevals{\cdot}{\cdot}{\cdot}$.
Applying projections during place evaluation ensures that every place
in the resulting borrow set reaches the reference's target, so borrow
sets record \emph{precise} aliasing information.
To keep the instrumentation minimal, borrow sets record only mutable
places.
Thus, \Rule{P-ShrRef} does not apply $\deref$ to places in $\bs_1$
(which also reach $\ell_1$), since
dereferencing them would produce shared (i.e., immutable) places.
At line~4, \Rule{I-Drop} invokes
\hyperref[fig:reference-invalidation]{$\mathsf{Invalidate}$} to
invalidate references borrowing \mcode{v} by setting them to
$\erefbot$.  This corresponds to
\hyperref[fig:abstract-evaluation-helpers]{$\mathsf{Invalidate}^\sharp$}
in the abstract interpreter.  However, since borrow sets are precise,
only references whose borrows actually conflict with the access are
invalidated.
Invalidation prevents \emph{out-of-order accesses}, where a borrowing
reference is used after a conflicting access to a place in its
borrow set.
In our semantics, such a use results in a borrow error.
At line~6, dereferencing \mcode{y} encounters an invalid reference, so
\Rule{P-Invalid} raises a borrow error, which eventually causes \Rule{Step-Bor-Err} to transition to
$\borrowerror$.

In this example, \mcode{y}'s type carries region \ticktt{y}, whose
loan set approximates \mcode{y}'s borrow set.  Recall that at line~3,
this region contains
$\rawset{\mcode{v}^{\mutx},\mcode{\deref\deref x}^{\texttt{s}}}$.  The
loan $\mcode{\deref\deref x}^{\texttt{s}}$ directly corresponds to
\mcode{\deref\deref x} in the borrow set, whereas $\mcode{v}^{\mutx}$
only approximates the borrow of \mcode{\deref v}.
This imprecision arises because $\Pevals{\cdot}{\cdot}{\cdot}$
propagates loans unchanged, whereas $\Peval{\cdot}{\cdot}$ applies
projections to borrowed places.
In general, for a reference place $p:\tref[r][\omega]{\tau}$, if
$\Peval{p}{\cstore}\Downarrow(\ell,\_)$, then the borrow set of the
reference stored at $\ell$ must be approximated by the region state
$\lmget{\astore}{r}$.  For soundness, every borrowed place must be
covered by a loan on that place or one of its prefixes, with at least
the same access restrictions (i.e., mutability).  This ensures that
abstract error detection covers concrete borrow errors.
This example shows how loan sets approximate borrow sets.  We now
explain what region equality approximates:
\begin{minted}[xleftmargin=1.5em,formatcom=\RustCodeLineNumbers,mathescape=true,escapeinside=||]{rust}
let mut v1 = 1; let mut v2 = Box::new(2); // $\mcode{v1}\mapsto1,\app \mcode{v2}\mapsto\ebox{2},\app \mcode{x}\mapsto\bot,\app \mcode{y}\mapsto\bot$ ($\textup{\Rule{I-Assign}}$)
let mut x: &'x mut i32 = &'r1 mut v1; // $\mcode{v1}\mapsto1,\app \mcode{v2}\mapsto\ebox{2},\app \mcode{x}\mapsto\emref{\mcode{v1}}{\rawset{\mcode{v1}}},\app \mcode{y}\mapsto\bot$ ($\textup{\Rule{I-Assign}}$)
let mut y: &'y1 mut &'y2 mut i32 = &'r2 mut x;
// $\mcode{v1}\mapsto1,\app \mcode{v2}\mapsto\ebox{2},\app \mcode{x}\mapsto\emref{\mcode{v1}}{\rawset{\mcode{v1}}},\app \mcode{y}\mapsto\emref{\mcode{x}}{\rawset{\mcode{x}}}$ ($\textup{\Rule{I-Assign}}$)
*y = &'r3 mut *v2; // $\mcode{v1}\mapsto1,\app \mcode{v2}\mapsto\ebox{2},\app \mcode{x}\mapsto\emref{\mcode{\deref v2}}{\rawset{\mcode{\deref v2}}},\app \mcode{y}\mapsto\emref{\mcode{x}}{\rawset{\mcode{x}}}$ ($\textup{\Rule{I-Assign}}$)
drop(v2); // $\mcode{v1}\mapsto1,\app \mcode{v2}\mapsto\bot,\app \mcode{x}\mapsto\erefbot,\app \mcode{y}\mapsto\emref{\mcode{x}}{\rawset{\mcode{x}}}$ ($\textup{\Rule{I-Drop}}$)
v2 = Box::new(3); // $\mcode{v1}\mapsto1,\app \mcode{v2}\mapsto\ebox{3},\app \mcode{x}\mapsto\erefbot,\app \mcode{y}\mapsto\emref{\mcode{x}}{\rawset{\mcode{x}}}$ ($\textup{\Rule{I-Assign}}$)
*x = 4; // borrow-error: x stores an invalid reference ($\textup{\Rule{P-Invalid}}$)
\end{minted}
At line~3, creating \mcode{y} makes \mcode{\deref y} and \mcode{x}
reach the same location.  At line~4, assigning through \mcode{\deref
  y} overwrites the reference stored in \mcode{x}, replacing its
borrow set with $\rawset{\mcode{\deref v2}}$.
Correspondingly, at line~3, the abstract interpreter equates
\ticktt{y2} and \ticktt{x}, giving
$[\ticktt{y2},\ticktt{x}]\mapsto \rawset{\mcode{v1}^{\mutx}}$.
At line~4, updating \ticktt{y2} also updates \ticktt{x}, giving
$[\ticktt{y2},\ticktt{x}]\mapsto
\rawset{\mcode{v1}^{\mutx},\mcode{\deref v2}^{\mutx}}$.
As in the previous example, the loan set approximates the concrete
borrow set.
Moreover, region equality approximates concrete aliasing.
If two reference places reach the same location, their regions must
belong to the same equivalence class.
This allows indirect updates to be reflected in all regions
representing that location.

Finally, we introduce function calls and the approximation
introduced by modular analysis:
\begin{minted}[xleftmargin=1.5em,formatcom=\RustCodeLineNumbers,mathescape=true,escapeinside=||]{rust}
// fn choose<'a,'b,'c>(a: &'a mut i32, b: &'b mut i32, d: bool) -> &'c mut i32
//  where 'a:'c, 'b:'c { if d { return a; } else { return b; } } // Correct region relations
let mut x = Box::new(1); let mut y = Box::new(2); // $\mcode{x}\mapsto\ebox{1},\app \mcode{y}\mapsto\ebox{2},\app \mcode{z}\mapsto\bot$ ($\textup{\Rule{I-Assign}}$)
let z: &'z mut i32 = choose(&'x mut *x, &'y mut *y, true);
// $\mcode{x}\mapsto\ebox{1},\app \mcode{y}\mapsto\ebox{2},\app \mcode{z}\mapsto\emref{\mcode{\deref x}}{\rawset{\mcode{\deref x}}}$ ($\textup{\Rule{Step-Call}, \Rule{Step-Return}}$); ['z]$\mapsto\rawset{\mcode{\deref x}^{\mutx},\mcode{\deref y}^{\mutx}}$ ($\textup{\Rule{I-Abs-Call}}$)
drop(x); // $\mcode{x}\mapsto\bot,\app \mcode{y}\mapsto\ebox{2},\app \mcode{z}\mapsto\erefbot$ ($\textup{\Rule{I-Drop}}$);
x = Box::new(2); // $\mcode{x}\mapsto\ebox{2},\app \mcode{y}\mapsto\ebox{2},\app \mcode{z}\mapsto\erefbot$ ($\textup{\Rule{I-Assign}}$)
println!("{}", *z); // borrow-error: z stores an invalid reference ($\textup{\Rule{P-Invalid}}$)
\end{minted}
\label{exm:fun-call1-concrete}
\Rule{Step-Call} enters the callee upon receiving a function call
signal from instruction evaluation.  We omit
the execution of \mcode{choose}.  Since the argument \mcode{d} is
true, \mcode{choose} returns
$\emref{\mcode{\deref x}}{\rawset{\mcode{\deref x}}}$.
\Rule{Step-Return} then assigns this reference to \mcode{z}.
In contrast, the abstract interpreter uses $\mathsf{TransferCall}$ to
apply the region relations in \mcode{choose}'s signature, propagating
the loans on both \mcode{\deref x} and \mcode{\deref y} to \ticktt{z},
i.e.,
$\ticktt{z}\mapsto\rawset{\mcode{\deref x}^{\mutx},\mcode{\deref
    y}^{\mutx}}$. Thus, the loan set of \ticktt{z} approximates the
concrete borrow set of \mcode{z}. This approximation is produced by
imprecision of the region relations in the function signature.

 \section{Correctness of Our Borrow Checking Algorithm}\label{sec:soundness-safety}

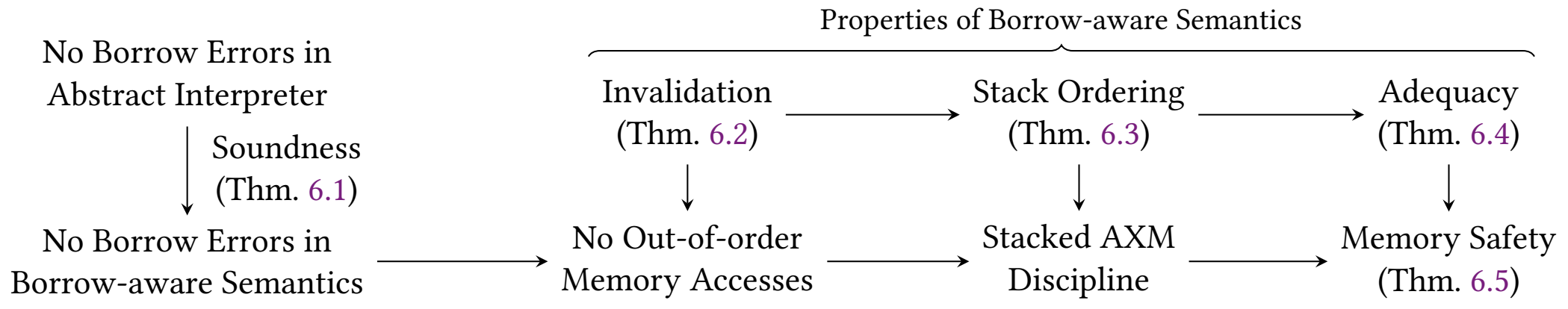
\begin{figure}[htbp]  
  \centering
  \resizebox{\linewidth}{!}{\begin{tikzpicture}[
      every node/.style={align=center, font=\small},
      >=stealth
    ]
      \node (abstract) at (0,0)
        {No Borrow Errors in\\Abstract Interpreter};
      \node (invalidation) at (4.6,-0.35)
        {Invalidation\\(\thmref{thm:main-reference-invalidation})};
      \node (orders) at (8.2,-0.35)
        {Stack Ordering\\(\thmref{def:main-borrow-invariant})};
      \node (adequacy) at (11.6,-0.35)
        {Adequacy\\(\thmref{thm:borrow-order-adequacy})};

      \node[fit=(invalidation)(orders)(adequacy), inner sep=0pt]
        (properties) {};
      \draw[decorate, decoration={brace, amplitude=3pt}]
        ($(properties.north west)+(0,2pt)$) --
        node[above=4pt, font=\footnotesize, inner sep=1pt]
          {Properties of Borrow-aware Semantics}
        ($(properties.north east)+(0,2pt)$);

      \coordinate (lower-row) at (0,-1.7);
      \node (concrete) at (abstract |- lower-row)
        {No Borrow Errors in\\Borrow-aware Semantics};
      \node (ordinary) at (invalidation |- lower-row)
        {No Out-of-order\\Memory Accesses};
      \node (stacked) at (orders |- lower-row)
        {Stacked AXM\\Discipline};
      \node (safety) at (adequacy |- lower-row)
        {Memory Safety\\(\thmref{thm:main-temporal-memory-safety})};

      \draw[->] (abstract) -- node[right=3pt]
        {Soundness\\(\thmref{thm:main-soundness})} (concrete);
      \draw[->] (invalidation) -- (orders);
      \draw[->] (orders) -- (adequacy);
      \draw[->] (invalidation) -- (ordinary);
      \draw[->] (orders) -- (stacked);
      \draw[->] (adequacy) -- (safety);
      \draw[->] (concrete) -- (ordinary);
      \draw[->] (ordinary) -- (stacked);
      \draw[->] (stacked) -- (safety);
    \end{tikzpicture}}
  \caption{Structure of the correctness proof for our borrow checking algorithm}
  \label{fig:proof-architecture}
\end{figure}

We follow \figref{fig:proof-architecture} to
introduce our correctness proof. It summarizes
how the soundness proof connects borrow checking to the stacked AXM
discipline,
and ultimately to memory safety.

\subsection{Soundness of the Abstract Interpreter}\label{ssec:soundness}

The soundness is formalized as a \emph{backward simulation} between the
abstract interpreter (\secref{sec:abstract-semantics}) and the
borrow-aware semantics (\secref{sec:concrete-semantics}). We define a
simulation relation, i.e., a \emph{concretization} relation, between
abstract and concrete states that characterizes how Polonius abstracts precise aliasing information
in the concrete semantics. We then show that it is preserved by each
step of the concrete semantics.
We write the concretization relation as
$\gamma:\hyperref[fig:abs-dom]{\Astore}\to
\hyperref[fig:program-state]{\Cstore}\to\prop$. 
As illustrated by the preceding examples, we
organize it into two components. The first relates region equality,
represented by union-find, to concrete aliasing relations.  The second
relates the map of region states to concrete memory states.
The latter involves three levels of correspondence: between abstract
states and concrete memory states, between region states and concrete
reference values, and between loan sets and borrow sets.  Modular
analysis further requires relating external loans to concrete borrow
sets from other function contexts. We leave the full formal definition
of $\gamma$ to the technical report.
\begin{theorem}[Soundness]
  \label{thm:main-soundness}
  Let $\prog$ be a \hyperref[fig:core-lang]{Core MIR} program that passes borrow checking, i.e.,
  $\BorrowCheck{\prog}=\mathsf{OK}(\mathcal A)$.  Let
  $\regstate{f}{n}{\cstore_1}{\kappa}$ be a reachable state of
  $\prog$, $i=g_f(n)$ be the current instruction and
  $\astore_1=A_f(n)$ be the abstract state computed at $n$.  If
  $\gamma(\astore_1,\cstore_1)$, then the following simulation
  properties hold, corresponding to the simulation diagrams
  in~\figref{fig:soundness-sim}:
  \begin{itemize}[wide=0pt]
    \item \textbf{Internal simulation.}
      $\Ieval{i}{\cstore_1}\Downarrow(\normalsig{n'},\cstore_2)
      \imply \exists\astore_2,\app
      \Ievals{i}{\astore_1}\Downarrow\astore_2
      \wedge\gamma(\astore_2,\cstore_2)$.

    \item \textbf{Function call and return.}
      $\regstate{f}{n}{\cstore_1}{\kappa}
      \xrightarrow{\text{\Rule{Step-Call}}}\mydots
      \xrightarrow{\text{\Rule{Step-Return}}}
      \regstate{f}{n'}{\cstore_2}{\kappa} \imply\exists\astore_2,\app
      \Ievals{i}{\astore_1}\Downarrow\astore_2 (\Rule{I-Abs-Call})
      \wedge\gamma(\astore_2,\cstore_2)$. Here, $\mydots$ denotes the
      concrete execution inside the callee.  In the actual proof, this
      execution is simulated by the abstract states computed for the
      callee.

    \item \textbf{Borrow-error simulation.}
      $\regstate{f}{n}{\cstore_1}{\kappa}\rightarrow\borrowerror
      \imply\Ievals{i}{\astore_1}\Uparrow\berrs$.
  \end{itemize}
\end{theorem}
By preserving the concretization relation during execution, we show
that the abstract semantics soundly approximates the concrete
semantics. Borrow-error simulation then ensures that abstract error
detection covers every concrete borrow error. Thus, programs that pass
borrow checking cannot reach a borrow-error state in the borrow-aware
semantics.
Establishing this simulation is highly nontrivial.
For \textbf{Internal simulation}, a representative difficulty is
assignment. The abstract interpreter propagates loans in a
type-directed manner through
$\hyperref[fig:abstract-loan-flow]{\mathsf{FlowLoans}}$, whereas the
concrete semantics performs ordinary value assignment.
Relating these operations requires reasoning about changes to the
entire concrete memory structure and showing that the concretization
relation is preserved.
\textbf{Function call and return} is the most challenging case.
As illustrated by the earlier example, the abstract interpreter
performs modular analysis and treats a call as a big-step operation
through $\hyperref[fig:abstract-call-and-init]{\mathsf{TransferCall}}$.
The concrete semantics instead enters the callee and executes its
body in small steps.
The key is to connect the callee's analysis with the caller's abstract
call operation. We show that
$\hyperref[fig:abstract-evaluation-rules]{\mathsf{CheckReturn}}$
ensures the signature's region relations (e.g., \ticktt{a}:\ticktt{c}
and \ticktt{b}:\ticktt{c} in \mcode{choose}) cover the callee's
possible loan flows. Combined with simulation within the callee, this
allows the caller to use
$\hyperref[fig:abstract-call-and-init]{\mathsf{TransferCall}}$ to
soundly approximate the entire concrete execution from call to return.

\subsection{Enforcement of the Stacked AXM Discipline in the
  Borrow-Aware Semantics}\label{ssec:stacked-axm}

Based on soundness, we formally prove the enforcement of Stacked AXM discipline in the borrow-aware
semantics. The basis of this proof is a \emph{borrow relation} between
aliased places:
\[
  p_2 \borarrow{\omega}_{\cstore} p_1 \iff \exists \bs,\;
  \Peval{p_2}{\cstore}\Downarrow(\_,\bs) \wedge
  p_1\in\bs\setminus\rawset{p_2} \wedge \omega=\mutmode{p_2}.
\]
Here, $\mutmode{\cdot}$ determines whether a place is mutable or
shared. For example, a variable storing a shared reference may itself
be mutable, whereas its dereference is a shared place.
The property of $\Peval{p_2}{\cstore}\Downarrow(\_,\bs)$ ensures that
all places in $\bs$ reach the same location as $p_2$. So
$p_1\in\bs\setminus\rawset{p_2}$ implies that $p_1$ and $p_2$ are
aliased. We can view $p_2\borarrow{\omega}_{\cstore}p_1$ as an order
over aliased places.
It is often the case that $p_2$ is a dereference $*p_2'$,
and in this case the reference $p_2'$ borrows (or points to) $p_1$.

Borrow orders and reference invalidation
(\figref{fig:reference-invalidation}) characterize \emph{out-of-order
  accesses}.  Before accessing a place, our semantics invalidates
references whose borrow sets conflict with the access (see
\Rule{I-Drop} and \Rule{I-Assign}).
That is, given $p_2\borarrow{\omega}_{\cstore}p_1$ , accessing $p_1$ invalidates $p_2$, making later evaluation of $p_2$ raise
borrow errors.  We formalize this property as follows:
\begin{theorem}[Out-of-Order Access Implies Borrow Error]
  \label{thm:main-reference-invalidation}
  Let $\regstate{f}{n}{\cstore}{\kappa}$ be a reachable state of a
  Core MIR program $\prog$.  Suppose
  $p_2\borarrow{\omega}_{\cstore}p_1$ and
  $\cstore'=\InvalidateRef{p_1}{\mathsf{acc}}{\cstore}$ (meaning
  that $p_1$ is accessed with $\mathsf{acc}$). If
  $\mathsf{acc}=\akwrite$, or $\mathsf{acc}=\akread$ and
  $\omega=\mathtt{mut}$, then $\Peval{p_2}{\cstore'}\Uparrow\berr$.
\end{theorem}
This theorem follows directly from the definitions of the borrow
relation and reference invalidation.
By this theorem, if borrow-aware semantics is free of borrow errors, then out-of-order access is not possible for a single layer of borrowing. By extending it to multiple layers of borrowing, stacked AXM is enforced.
We observe from~\thmref{thm:main-reference-invalidation} that creating
a new borrow can only extend the existing borrow order among valid
places, without splitting it.  By transitivity of the borrow order, we
can prove that \emph{any} two aliased places are ordered whenever at
least one of them is mutable:
\begin{theorem}[Aliased Places Are Ordered by Borrow Relation]
  \label{def:main-borrow-invariant}
  Let $\prog$ be a Core MIR program and
  $\regstate{f}{n}{\cstore}{\kappa}$ be a reachable state of $\prog$.
  For any two distinct places $p_1$ and $p_2$, if
  $\Peval{p_1}{\cstore}\Downarrow(\ell,\_)$ and
  $\Peval{p_2}{\cstore}\Downarrow(\ell,\_)$ for some $\ell$, then:
\begin{enumerate}[leftmargin=*]
    \item If both places are mutable, then either
      $p_2\borarrow{\mathtt{mut}}_{\cstore}p_1$ or
      $p_1\borarrow{\mathtt{mut}}_{\cstore}p_2$ holds.
    \item If $p_1$ is shared and $p_2$ is mutable, then
      $p_1\borarrow{\mathtt{shr}}_{\cstore}p_2$ holds.
  \end{enumerate}
\end{theorem}

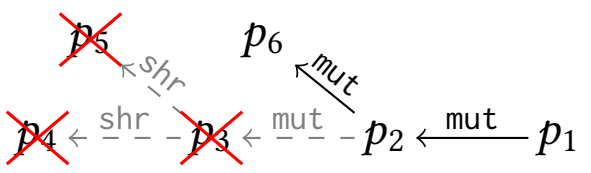
\begin{wrapfigure}{r}{0.32\textwidth}
  \centering
  \begin{tikzpicture}[
    every node/.style={font=\small, inner sep=1.5pt},
    >=to
  ]
    \node (p4) at (0,0) {$p_4$};
    \node (p3) at (1.15,0) {$p_3$};
    \node (p2) at (2.3,0) {$p_2$};
    \node (p1) at (3.45,0) {$p_1$};
    \node (p5) at (0.35,0.65) {$p_5$};
    \node (p6) at (1.5,0.65) {$p_6$};
    \draw[<-, gray, dashed] (p4) --
      node[above] {$\scriptstyle\mathtt{shr}$} (p3);
    \draw[<-, gray, dashed] (p3) --
      node[above] {$\scriptstyle\mathtt{mut}$} (p2);
    \draw[<-] (p2) --
      node[above] {$\scriptstyle\mathtt{mut}$} (p1);
    \draw[<-, gray, dashed] (p5) --
      node[above, sloped] {$\scriptstyle\mathtt{shr}$} (p3);
    \draw[<-] (p6) --
      node[above, sloped] {$\scriptstyle\mathtt{mut}$} (p2);
    \foreach \p in {p3,p4,p5} {
      \draw[red, line width=0.6pt]
        (\p.north west) -- (\p.south east)
        (\p.south west) -- (\p.north east);
    }
  \end{tikzpicture}
  \caption{A new mutable borrow $p_6$ of $p_2$ invalidates
    $p_3$, $p_4$, and $p_5$.}
  \label{fig:borrow-order-invalidation}
\end{wrapfigure}
A key step in proving this theorem is showing that creating a new
reference preserves the total order among mutable places. As shown
in~\figref{fig:borrow-order-invalidation}, a new reference may create
a branch that makes two mutable places ($p_6$ and $p_3$) incomparable.
However, \thmref{thm:main-reference-invalidation} ensures that places
on the old branch are invalidated. Since
\thmref{def:main-borrow-invariant} only concerns valid places (whose
evaluations do not raise borrow errors), all valid mutable places are
still ordered.

\thmref{def:main-borrow-invariant}
ensures that all mutable aliased places are ordered like a stack (i.e., the borrow chain as discussed in~\secref{sec:overview}), while shared places
form tree-like branches above it, with no required ordering among
them. Because borrow orders are transitive, any two distinct mutable and aliased places are ordered. When borrow errors are absent in borrow-aware semantics, applying the contrapositive of \thmref{thm:main-reference-invalidation} to this ordering guarantees \emph{stacked AXM discipline}, i.e., no out-of-order accesses to places on the stack (and tree branches for shared places) are possible.

\subsection{From Stacked AXM Discipline to Temporal Memory Safety}\label{ssec:memory-safety}

In type checked safe Rust, temporal memory errors could only be caused by accessing dangling pointers to deallocated
memory.  To prove that the stacked AXM discipline implies absence of such memory errors, we prove that accessing dangling pointers could only happen for out-of-order accesses to the above borrow stack (tree), hence conflicting with stacked AXM.
The following \emph{adequacy property} connects the borrow order to operations that may deallocate a reference's target.
\begin{theorem}[Adequacy of the Borrow Order]
  \label{thm:borrow-order-adequacy}
  Let $\regstate{f}{n}{\cstore}{\kappa}$ be a reachable state of a
  program $\prog$.  Consider distinct places $p_1=q[p_0]$ and $p_2$,
  where $q\neq[\cdot]$ and $\deref^r\notin q$.  If
  $\Peval{p_0}{\cstore}\Downarrow(\ell,\_)$ and
  $\Vread{p_2}{\cstore}{\eoref{q[\ell]}{\bs}}$ (i.e., $p_2$ must be a
  move path), then $\deref p_2\borarrow{\omega}_{\cstore}p_1$, i.e., $p_1\in\bs$.
\end{theorem}
This property states that the borrow order captures all borrow
relations that could potentially cause temporal memory errors.
Here, $q$ contains only field projections and Box dereferences.
Dropping (freeing) $p_0$ releases the value at $\ell$ and its owned subvalues,
including the value at $q[\ell]$, which is pointed to by the reference
stored in $p_2$.  This could and is the only way to leave a dangling reference $p_2$ and cause a memory error by accessing $p_2$ (as shown by examples
in~\secref{sec:concrete-semantics}).
In this case, the adequacy property requires
$\deref p_2\borarrow{\omega}_{\cstore}p_1$.
The stacked AXM discipline then ensures that dropping $p_0$ (prefix of $p_1$) could not happen before accessing $p_2$ because this is an out-of-order access.

To prove memory safety, we extend the borrow-aware semantics with basic memory allocation and
deallocation for modeling errors caused by accessing freed
memory. Each reference records its target memory block when created,
and the memory state tracks which blocks are allocated or freed. A
reference is \emph{dangling} if the block in its recorded address has
been freed.
\begin{theorem}[Borrow Checking Ensures Temporal Memory Safety]
  \label{thm:main-temporal-memory-safety}
  Let $\prog$ be a Core MIR program that passes borrow checking, i.e.,
  $\BorrowCheck{\prog}=\mathsf{OK}(\mathcal A)$.  For any reachable
  state $s$ of $\prog$, its next step cannot access a reference
  whose recorded memory block has been freed.
\end{theorem}
The proof establishes a backward simulation from the extended
semantics to the original borrow-aware semantics.  The adequacy
property and reference invalidation ensure that references are
invalidated before their target blocks are freed, so any temporal
memory error implies a borrow error.  Soundness excludes such borrow (memory)
errors in programs that pass borrow
checking.

\section{Rocq Formalization and Evaluation}\label{sec:eval}

\tabref{tab:rocq-loc} reports the lines of code of our Rocq
development. We manually implemented the abstract interpreter and
defined the borrow-aware semantics using CompCert's dataflow
analysis~\cite{leroy09,Leroy-backend,compcert-va}.
The dataflow framework provides lattice libraries and a solver based
on Kildall's algorithm~\cite{kildall}, which we use to implement the
abstract domains and the solver in~\algoref{alg:borrow-check}.
For the correctness
\begin{wraptable}{r}{0.4\textwidth}
  \centering
  \small
  \caption{Size of the development}
  \label{tab:rocq-loc}
  \begin{tabular}{l r}
    \toprule
    \textbf{Component} & \textbf{Lines} \\
    \midrule
    Abstract interpreter & 4,369\\
    Borrow-aware semantics & 1,454\\
    Correctness proof (generated) & \\
    \quad Soundness & 16,965\\
    \quad Stacked AXM and adequacy & 38,740\\
    \quad Memory safety & 5,232\\
    \midrule
    \textbf{Total Rocq} & 66,760\\
    \bottomrule
  \end{tabular}
\end{wraptable}proof, we first developed a complete paper proof spanning roughly 100
pages, which is included in the technical report. We manually wrote
all definitions, theorem statements, and proof outlines, and then used
AI tools to refine the proofs by explicitly stating the goals to be
proved, listing the assumptions, and detailing each derivation step
following the proof outlines.
This required substantial conceptual work, including the proof
structure shown in~\figref{fig:proof-architecture}.
We then used Codex~\cite{codex} with GPT-6 Astra to generate the Rocq
proofs. This took about two weeks and went relatively smoothly, as
most conceptual difficulties had already been resolved on paper.
Our experience shows that LLMs can support large-scale verification
when guided by a clear proof structure. The paper proof allows us to
focus on key arguments and establish a workable proof strategy without
getting involved in tedious proof details. It also helps reduce errors
during proof generation.

To evaluate whether our formalized borrow checker agrees with
\mcode{rustc}'s borrow checker, we implement a frontend that
translates Rust source programs into our core MIR. We then test the
borrow checker on 510 Rust programs drawn primarily from the
\mcode{borrowck} and \mcode{nll} suites of
\mcode{rustc}~\cite{rustc-tests} and the Polonius test
suite~\cite{rust-polonius}.  We exclude tests that depend on
unsupported features such as traits, closures, and unsafe code.
We compare our results with those of \mcode{rustc 1.101.0-nightly} using
\scode{-Zpolonius=next} to enable the Polonius
checker~\cite{polonius-alpha}.
Our checker agrees with \mcode{rustc}'s Polonius checker on 499 of the
510 programs (97.8\%).  Among the remaining 11 programs, four are
accepted by Polonius but rejected by our checker, mainly due to
unsupported features such as two-phase borrow~\cite{two-phase-borrow}
and smarter checking for dropping a \mcode{Box}~\cite{rustc-drop-box}.
The other seven are accepted by our checker but rejected by Polonius,
mostly because \mcode{rustc} imposes stricter checks on region
parameters in function signatures. Further details of the experiments
are available in the submitted~code.

 \section{Related Work and Conclusion}\label{sec:related}

\begingroup

The success of Rust has stimulated growing research interest in borrow
checking (see~\tabref{tab:related-work}).  Many
works study the underlying principles of borrowing independently of
the algorithms in \mcode{rustc}, often through type systems or symbolic execution.
We discuss them below.

\myparagraph{RustBelt} RustBelt~\cite{RustBelt} formalizes borrow
checking as an ownership-based type system. It provides a semantic interpretation of the type system using the
Iris framework~\cite{iris15,iris-ground}, interpreting the typing
rules as rules in separation logic. Borrowing is modeled through a
custom logic called \emph{lifetime logic}, which defines a dynamic
mechanism for transferring resources and specifying when they must be
returned. We refer to this approach as modeling borrowing through
\emph{resource transfer}.
RustBelt targets a language called $\lambda_{\text{rust}}$ that is
different from MIR, and the type checking on $\lambda_{\text{rust}}$
still relies on explicit lifetime annotations provided by
\mcode{rustc}'s borrow checking.

\myparagraph{Aeneas} Aeneas~\cite{Aeneas,sound-borrow} also adopts a
resource transfer view of borrowing. It defines a semantics called
LLBC to describe how \emph{values} are temporarily moved and later
returned through borrowing. From this semantics, Aeneas develops a
borrow checker based on symbolic execution. LLBC uses \emph{loan
  identifiers} to track the transfer and return of values. For example, the loan
identifier $\ell$ below records that \mcode{p} borrows the value of
\mcode{v} and allows the value to be returned when the borrow ends:
\begin{minted}[xleftmargin=1.5em,formatcom=\RustCodeLineNumbers,mathescape=true,escapeinside=||]{rust}
let mut v = 1;      // LLBC: $\mcode{v}\mapsto 1$; Ours: $\mcode{v}\mapsto 1$
let mut p = &mut v; // LLBC: $\mcode{v}\mapsto \mathsf{loan}^\mutx\app \ell, \app \mcode{p}\mapsto \mathsf{borrow}^\mutx\app \ell\app 1$; Ours: $\tt \mcode{v}\mapsto 1,\app \mcode{p}\mapsto \emref{v}{\rawset{v}}$
v = 2; // Borrow ends; LLBC: $\mcode{v}\mapsto 2, \app \mcode{p}\mapsto \bot$; Ours: $\tt \mcode{v}\mapsto 2,\app \mcode{p}\mapsto \erefbot$
\end{minted}
Our borrow-aware semantics uses a structured memory model similar to
that of LLBC, but uses borrow sets instead of loan identifiers in
references to track borrowing. We show that this change is key to
providing a semantic interpretation of Polonius, as recovering borrow
sets from loan identifiers is difficult.
The work on LLBC's soundness~\cite{sound-borrow} further presents a
paper proof of a simulation between LLBC and a low-level semantics
with a CompCert-based memory model. There is also ongoing work~\cite{sound-borrow-formalize} on
formalizing LLBC and its simulation proofs in Rocq.

\begin{table}[t]
  \smaller
  \caption{Description and comparison of existing work on static and
    dynamic borrow checking for Rust.}\label{tab:related-work}
  \centering
\begin{tabular}{l | c|c|c|c}
\midrule
    \begin{tabular}[c]{@{}l@{}}Work on\\Borrow Checking\end{tabular} &
    \begin{tabular}[c]{@{}c@{}}Model \mcode{rustc}'s\\Algorithms\end{tabular} &
    \begin{tabular}[c]{@{}c@{}}Interpretation\\of Borrowing\end{tabular} &
\begin{tabular}[c]{@{}c@{}}Techniques of\\Program Analysis\end{tabular} &
\begin{tabular}[c]{@{}c@{}}Program Property\\to Check/Prove\end{tabular}
    \\
    \midrule
    \rowcolor{gray!15}
    RustBelt \cite{RustBelt}
    & \xmark & Resource Transfer
    & Type System& Semantic Type Safety \\

    Aeneas \cite{Aeneas,sound-borrow}
    & \xmark & Resource Transfer
    & Symbolic Execution & Safe Dynamic Borrowing \\

    \rowcolor{gray!15}
    Oxide \cite{weiss2019oxide}
    & \xmark & Aliasing & Type System&  Syntactic Type Safety \\

\begin{tabular}[c]{@{}l@{}}Stacked / Tree\\Borrows \cite{stacked-borrow,tree-borrow}\end{tabular}
    & \xmark & Aliasing &  Dynamic Testing
    & \begin{tabular}[c]{@{}c@{}}Stacked/Tree\\AXM Discipline\end{tabular} \\

\rowcolor{gray!15}
    This Work
    & \cmark & Aliasing & Abstract Interpretation
    & \begin{tabular}[c]{@{}c@{}}Stacked AXM Discipline\end{tabular}\\
  \end{tabular}
\end{table}

\myparagraph{Oxide} Oxide~\cite{weiss2019oxide} proposes an
alias-based type system for borrow checking, adopting a view similar
to Polonius.
Unlike our work, Oxide aims to provide a rich type system for surface
Rust rather than directly modeling Polonius's analysis.
For example, Oxide does not model Polonius's region equality for
nested references, region parameters in composite data types, or heap
allocation through \mcode{Box}. However, it supports higher-order
functions, polymorphism, and move checking, which we do not cover.
For soundness, Oxide proves type safety through preservation and
progress over a conventional operational semantics that does not
model the stacked AXM discipline.

\myparagraph{Stacked/Tree Borrows} Stacked
Borrows~\cite{stacked-borrow} defines a low-level semantics that uses
pointer provenance to distinguish aliased pointers and maintains a
borrow stack at each memory location to enforce the stacked AXM
discipline.  Tree Borrows~\cite{tree-borrow} relaxes several
restrictions of Stacked Borrows by replacing borrow stacks with trees,
accepting more borrowing patterns.
These works primarily focus on detecting undefined behavior arising
from unsafe code~\cite{Miri}, which is orthogonal to our goal of
verifying borrow checking.
We show that our borrow-aware semantics enforces a similar stacked AXM
discipline, with the stack represented implicitly by the borrow order.
This representation may also extend to tree structures. Establishing a
simulation between our borrow-aware semantics and Stacked Borrows or
Tree Borrows is a direction for future work.

\myparagraph{Formalization of Polonius}
Polonius was originally implemented in
Datalog~\cite{rust-polonius,Polonius-thesis}, using constraints
generated by \mcode{rustc}'s NLL analysis.
The Polonius developers also explored a forward-only formulation
without liveness analysis~\cite{forward-polonius}.  Unlike their
Datalog formulation, we implement forward-only Polonius as an abstract
interpreter with a nontrivial abstract domain and transfer functions.
The a-mir-formality project~\cite{a-mir-formality} formalizes MIR and
models NLL and Polonius through inference rules written in Rust,
aiming to provide a specification of borrow checking in \mcode{rustc}.

\myparagraph{Other work on borrow checking}
Featherweight Rust~\cite{light-rust} is an early formalization of
Rust's type system. It uses lexical lifetimes rather than the
non-lexical lifetimes of NLL and Polonius.  Its language excludes
features such as function calls and enums.
The Move borrow checker~\cite{move-borrow-checker} uses regular
expressions to track reference reachability in the Move
language~\cite{blackshear2019move}.  Move excludes features such as
nested references and references stored in structs or enums.  Under
these restrictions, its memory-safety guarantee does not require the
region-based analysis used by Polonius.

\myparagraph{Work on Rust verification}
Many Rust verification tools, including Verus~\cite{Verus},
Prusti~\cite{Prusti,PCG}, Creusot~\cite{Creusot},
RustHorn~\cite{RustHorn}, Flux~\cite{Flux}, and Thrust~\cite{Thrust}, use the guarantees
provided by borrow checking to simplify memory reasoning and focus on
functional correctness.  We believe that our verified borrow checker
can provide a formal foundation for this approach.

\myparagraph{Conclusion}
We have presented the first comprehensive verified implementation of the Polonius borrow checking algorithm for Rust. We observe that the essence of Polonius could be described via abstract interpretation, hence make its verification feasible. In the future, we plan to integrate this borrow checker into CompCert to build a fully verified compiler for a subset of Rust.

\par\endgroup

 \par
\endgroup

\section*{Data-Availability Statement}

The technical report and Rocq formalization are available in the
supplementary material. The technical report includes the full paper
formalization and proofs. The Rocq formalization includes the borrow
checker's source code and build scripts, the machine-checked
correctness proof, and conformance tests. This paper is under
submission and we will open source the code and technical reports
after the submission.


\begin{thebibliography}{50}



\ifx \showCODEN    \undefined \def \showCODEN     #1{\unskip}     \fi
\ifx \showISBNx    \undefined \def \showISBNx     #1{\unskip}     \fi
\ifx \showISBNxiii \undefined \def \showISBNxiii  #1{\unskip}     \fi
\ifx \showISSN     \undefined \def \showISSN      #1{\unskip}     \fi
\ifx \showLCCN     \undefined \def \showLCCN      #1{\unskip}     \fi
\ifx \shownote     \undefined \def \shownote      #1{#1}          \fi
\ifx \showarticletitle \undefined \def \showarticletitle #1{#1}   \fi
\ifx \showURL      \undefined \def \showURL       {\relax}        \fi
\providecommand\bibfield[2]{#2}
\providecommand\bibinfo[2]{#2}
\providecommand\natexlab[1]{#1}
\providecommand\showeprint[2][]{arXiv:#2}
\makeatletter
\@ifundefined{NAT@parse@date}{}{\let\NAT@parse@date@orig\NAT@parse@date}
\@ifundefined{NAT@parse@date}{}{\def\NAT@parse@date#1#2#3#4#5#6@@{\NAT@parse@date@orig#1#2#3#4#5#6@@\def\NAT@tempyear{0000}\def\NAT@tempexlab{{?}}\ifx\NAT@year\NAT@tempyear\ifx\NAT@exlab\NAT@tempexlab\def\NAT@date{[n.\,d.]}\else\edef\NAT@date{[n.\,d.]\NAT@exlab}\fi\fi}}
\makeatother

\bibitem[Andersen(1994)]{andersen1994program}
\bibfield{author}{\bibinfo{person}{Lars~Ole Andersen}.}
  \bibinfo{year}{1994}\natexlab{}.
\newblock \emph{\bibinfo{title}{Program analysis and specialization for the C
  programming language}}.
\newblock \bibinfo{thesistype}{Ph.\,D. Dissertation}.
  \bibinfo{school}{University of Copenhagen}.
\newblock
\urldef\tempurl \url{https://www.cs.cornell.edu/courses/cs711/2005fa/papers/andersen-thesis94.pdf}
\showURL{\tempurl}


\bibitem[Astrauskas et~al\mbox{.}(2019)]{Prusti}
\bibfield{author}{\bibinfo{person}{Vytautas Astrauskas}, \bibinfo{person}{Peter
  M\"{u}ller}, \bibinfo{person}{Federico Poli}, {and}
  \bibinfo{person}{Alexander~J. Summers}.} \bibinfo{year}{2019}\natexlab{}.
\newblock \showarticletitle{Leveraging rust types for modular specification and
  verification}.
\newblock \bibinfo{journal}{\emph{Proc. ACM Program. Lang.}}
  \bibinfo{volume}{3}, \bibinfo{number}{OOPSLA}, Article
  \bibinfo{articleno}{147} (\bibinfo{date}{Oct.} \bibinfo{year}{2019}),
  \bibinfo{numpages}{30}~pages.
\newblock
\href{https://doi.org/10.1145/3360573}{doi:\nolinkurl{10.1145/3360573}}


\bibitem[Blackshear et~al\mbox{.}(2019)]{blackshear2019move}
\bibfield{author}{\bibinfo{person}{Sam Blackshear}, \bibinfo{person}{Evan
  Cheng}, \bibinfo{person}{David~L Dill}, \bibinfo{person}{Victor Gao},
  \bibinfo{person}{Ben Maurer}, \bibinfo{person}{Todd Nowacki},
  \bibinfo{person}{Alistair Pott}, \bibinfo{person}{Shaz Qadeer},
  \bibinfo{person}{Dario~Russi Rain}, \bibinfo{person}{Stephane Sezer},
  {et~al\mbox{.}}} \bibinfo{year}{2019}\natexlab{}.
\newblock \showarticletitle{Move: A language with programmable resources}.
\newblock \bibinfo{journal}{\emph{Libra Assoc}}  \bibinfo{volume}{1}
  (\bibinfo{year}{2019}).
\newblock
\urldef\tempurl \url{https://developers.diem.com/papers/diem-move-a-language-with-programmable-resources/2019-09-26.pdf}
\showURL{\tempurl}


\bibitem[Blazy et~al\mbox{.}(2013)]{compcert-va}
\bibfield{author}{\bibinfo{person}{Sandrine Blazy}, \bibinfo{person}{Vincent
  Laporte}, \bibinfo{person}{Andr{\'e} Maroneze}, {and} \bibinfo{person}{David
  Pichardie}.} \bibinfo{year}{2013}\natexlab{}.
\newblock \showarticletitle{Formal Verification of a C Value Analysis Based on
  Abstract Interpretation}. In \bibinfo{booktitle}{\emph{Static Analysis}},
  \bibfield{editor}{\bibinfo{person}{Francesco Logozzo} {and}
  \bibinfo{person}{Manuel F{\"a}hndrich}} (Eds.). \bibinfo{publisher}{Springer
  Berlin Heidelberg}, \bibinfo{address}{Berlin, Heidelberg},
  \bibinfo{pages}{324--344}.
\newblock
\showISBNx{978-3-642-38856-9}


\bibitem[Crichton et~al\mbox{.}(2023)]{will-conceptual}
\bibfield{author}{\bibinfo{person}{Will Crichton}, \bibinfo{person}{Gavin
  Gray}, {and} \bibinfo{person}{Shriram Krishnamurthi}.}
  \bibinfo{year}{2023}\natexlab{}.
\newblock \showarticletitle{A Grounded Conceptual Model for Ownership Types in
  Rust}.
\newblock \bibinfo{journal}{\emph{Proc. ACM Program. Lang.}}
  \bibinfo{volume}{7}, \bibinfo{number}{OOPSLA2}, Article
  \bibinfo{articleno}{265} (\bibinfo{date}{Oct.} \bibinfo{year}{2023}),
  \bibinfo{numpages}{29}~pages.
\newblock
\href{https://doi.org/10.1145/3622841}{doi:\nolinkurl{10.1145/3622841}}


\bibitem[Denis et~al\mbox{.}(2022)]{Creusot}
\bibfield{author}{\bibinfo{person}{Xavier Denis},
  \bibinfo{person}{Jacques-Henri Jourdan}, {and} \bibinfo{person}{Claude
  March\'{e}}.} \bibinfo{year}{2022}\natexlab{}.
\newblock \showarticletitle{Creusot: A Foundry for the Deductive Verification
  of Rust Programs}. In \bibinfo{booktitle}{\emph{Formal Methods and Software
  Engineering: 23rd International Conference on Formal Engineering Methods,
  ICFEM 2022, Madrid, Spain, October 24–27, 2022, Proceedings}} (Madrid,
  Spain). \bibinfo{publisher}{Springer-Verlag}, \bibinfo{address}{Berlin,
  Heidelberg}, \bibinfo{pages}{90–105}.
\newblock
\showISBNx{978-3-031-17243-4}
\href{https://doi.org/10.1007/978-3-031-17244-1_6}{doi:\nolinkurl{10.1007/978-3-031-17244-1_6}}


\bibitem[Grannan et~al\mbox{.}(2025)]{PCG}
\bibfield{author}{\bibinfo{person}{Zachary Grannan}, \bibinfo{person}{Aurel
  B{\'i}l{\'y}}, \bibinfo{person}{Jon{\'a}{\v s} Fiala},
  \bibinfo{person}{Jasper Geer}, \bibinfo{person}{Markus de Medeiros},
  \bibinfo{person}{Peter M{\"u}ller}, {and} \bibinfo{person}{Alexander~J.
  Summers}.} \bibinfo{year}{2025}\natexlab{}.
\newblock \showarticletitle{Place Capability Graphs: A General-Purpose Model of
  Rust’s Ownership and Borrowing Guarantees}.
\newblock \bibinfo{journal}{\emph{Proc. ACM Program. Lang.}}
  \bibinfo{volume}{9}, \bibinfo{number}{OOPSLA2}, Article
  \bibinfo{articleno}{344} (\bibinfo{date}{Oct.} \bibinfo{year}{2025}),
  \bibinfo{numpages}{28}~pages.
\newblock
\href{https://doi.org/10.1145/3763122}{doi:\nolinkurl{10.1145/3763122}}


\bibitem[Hind(2001)]{pointer-survey}
\bibfield{author}{\bibinfo{person}{Michael Hind}.}
  \bibinfo{year}{2001}\natexlab{}.
\newblock \showarticletitle{Pointer analysis: haven't we solved this problem
  yet?}. In \bibinfo{booktitle}{\emph{Proceedings of the 2001 ACM
  SIGPLAN-SIGSOFT Workshop on Program Analysis for Software Tools and
  Engineering}} (Snowbird, Utah, USA) \emph{(\bibinfo{series}{PASTE '01})}.
  \bibinfo{publisher}{Association for Computing Machinery},
  \bibinfo{address}{New York, NY, USA}, \bibinfo{pages}{54–61}.
\newblock
\showISBNx{1581134134}
\href{https://doi.org/10.1145/379605.379665}{doi:\nolinkurl{10.1145/379605.379665}}


\bibitem[Ho et~al\mbox{.}(2024)]{sound-borrow}
\bibfield{author}{\bibinfo{person}{Son Ho}, \bibinfo{person}{Aymeric Fromherz},
  {and} \bibinfo{person}{Jonathan Protzenko}.} \bibinfo{year}{2024}\natexlab{}.
\newblock \showarticletitle{Sound Borrow-Checking for Rust via Symbolic
  Semantics}.
\newblock \bibinfo{journal}{\emph{Proc. ACM Program. Lang.}}
  \bibinfo{volume}{8}, \bibinfo{number}{ICFP}, Article \bibinfo{articleno}{251}
  (\bibinfo{date}{Aug.} \bibinfo{year}{2024}), \bibinfo{numpages}{29}~pages.
\newblock
\href{https://doi.org/10.1145/3674640}{doi:\nolinkurl{10.1145/3674640}}


\bibitem[Ho and Protzenko(2022)]{Aeneas}
\bibfield{author}{\bibinfo{person}{Son Ho} {and} \bibinfo{person}{Jonathan
  Protzenko}.} \bibinfo{year}{2022}\natexlab{}.
\newblock \showarticletitle{Aeneas: Rust verification by functional
  translation}.
\newblock \bibinfo{journal}{\emph{Proc. ACM Program. Lang.}}
  \bibinfo{volume}{6}, \bibinfo{number}{ICFP}, Article \bibinfo{articleno}{116}
  (\bibinfo{date}{Aug.} \bibinfo{year}{2022}), \bibinfo{numpages}{31}~pages.
\newblock
\href{https://doi.org/10.1145/3547647}{doi:\nolinkurl{10.1145/3547647}}


\bibitem[Huey(2026)]{polonius-alpha}
\bibfield{author}{\bibinfo{person}{Jack Huey}.}
  \bibinfo{year}{2026}\natexlab{}.
\newblock \bibinfo{title}{Enabling the Next Iteration of the Borrow Checker on
  Nightly}.
\newblock
\urldef\tempurl \url{https://blog.rust-lang.org/2026/08/04/enabling-polonius-alpha-on-nightly/}
\showURL{\tempurl}


\bibitem[Jourdan et~al\mbox{.}(2015)]{verified-static-analyzer}
\bibfield{author}{\bibinfo{person}{Jacques-Henri Jourdan},
  \bibinfo{person}{Vincent Laporte}, \bibinfo{person}{Sandrine Blazy},
  \bibinfo{person}{Xavier Leroy}, {and} \bibinfo{person}{David Pichardie}.}
  \bibinfo{year}{2015}\natexlab{}.
\newblock \showarticletitle{A Formally-Verified C Static Analyzer}. In
  \bibinfo{booktitle}{\emph{Proceedings of the 42nd Annual ACM SIGPLAN-SIGACT
  Symposium on Principles of Programming Languages}} (Mumbai, India)
  \emph{(\bibinfo{series}{POPL '15})}. \bibinfo{publisher}{Association for
  Computing Machinery}, \bibinfo{address}{New York, NY, USA},
  \bibinfo{pages}{247–259}.
\newblock
\showISBNx{9781450333009}
\href{https://doi.org/10.1145/2676726.2676966}{doi:\nolinkurl{10.1145/2676726.2676966}}


\bibitem[Jung et~al\mbox{.}(2019)]{stacked-borrow}
\bibfield{author}{\bibinfo{person}{Ralf Jung}, \bibinfo{person}{Hoang-Hai
  Dang}, \bibinfo{person}{Jeehoon Kang}, {and} \bibinfo{person}{Derek Dreyer}.}
  \bibinfo{year}{2019}\natexlab{}.
\newblock \showarticletitle{Stacked borrows: an aliasing model for Rust}.
\newblock \bibinfo{journal}{\emph{Proc. ACM Program. Lang.}}
  \bibinfo{volume}{4}, \bibinfo{number}{POPL}, Article \bibinfo{articleno}{41}
  (\bibinfo{date}{Dec.} \bibinfo{year}{2019}), \bibinfo{numpages}{32}~pages.
\newblock
\href{https://doi.org/10.1145/3371109}{doi:\nolinkurl{10.1145/3371109}}


\bibitem[Jung et~al\mbox{.}(2017)]{RustBelt}
\bibfield{author}{\bibinfo{person}{Ralf Jung}, \bibinfo{person}{Jacques-Henri
  Jourdan}, \bibinfo{person}{Robbert Krebbers}, {and} \bibinfo{person}{Derek
  Dreyer}.} \bibinfo{year}{2017}\natexlab{}.
\newblock \showarticletitle{RustBelt: securing the foundations of the Rust
  programming language}.
\newblock \bibinfo{journal}{\emph{Proc. ACM Program. Lang.}}
  \bibinfo{volume}{2}, \bibinfo{number}{POPL}, Article \bibinfo{articleno}{66}
  (\bibinfo{date}{Dec.} \bibinfo{year}{2017}), \bibinfo{numpages}{34}~pages.
\newblock
\href{https://doi.org/10.1145/3158154}{doi:\nolinkurl{10.1145/3158154}}


\bibitem[Jung et~al\mbox{.}(2026)]{Miri}
\bibfield{author}{\bibinfo{person}{Ralf Jung}, \bibinfo{person}{Benjamin
  Kimock}, \bibinfo{person}{Christian Poveda},
  \bibinfo{person}{Eduardo~S\'{a}nchez Mu\~{n}oz}, \bibinfo{person}{Oli
  Scherer}, {and} \bibinfo{person}{Qian Wang}.}
  \bibinfo{year}{2026}\natexlab{}.
\newblock \showarticletitle{Miri: Practical Undefined Behavior Detection for
  Rust}.
\newblock \bibinfo{journal}{\emph{Proc. ACM Program. Lang.}}
  \bibinfo{volume}{10}, \bibinfo{number}{POPL}, Article \bibinfo{articleno}{48}
  (\bibinfo{date}{Jan.} \bibinfo{year}{2026}), \bibinfo{numpages}{29}~pages.
\newblock
\href{https://doi.org/10.1145/3776690}{doi:\nolinkurl{10.1145/3776690}}


\bibitem[JUNG et~al\mbox{.}(2018)]{iris-ground}
\bibfield{author}{\bibinfo{person}{RALF JUNG}, \bibinfo{person}{ROBBERT
  KREBBERS}, \bibinfo{person}{JACQUES-HENRI JOURDAN}, \bibinfo{person}{ALEŠ
  BIZJAK}, \bibinfo{person}{LARS BIRKEDAL}, {and} \bibinfo{person}{DEREK
  DREYER}.} \bibinfo{year}{2018}\natexlab{}.
\newblock \showarticletitle{Iris from the ground up: A modular foundation for
  higher-order concurrent separation logic}.
\newblock \bibinfo{journal}{\emph{Journal of Functional Programming}}
  \bibinfo{volume}{28} (\bibinfo{year}{2018}), \bibinfo{pages}{e20}.
\newblock
\href{https://doi.org/10.1017/S0956796818000151}{doi:\nolinkurl{10.1017/S0956796818000151}}


\bibitem[Jung et~al\mbox{.}(2015)]{iris15}
\bibfield{author}{\bibinfo{person}{Ralf Jung}, \bibinfo{person}{David Swasey},
  \bibinfo{person}{Filip Sieczkowski}, \bibinfo{person}{Ksper Svendsen},
  \bibinfo{person}{Aaron Turon}, \bibinfo{person}{Lars Birkedal}, {and}
  \bibinfo{person}{Derek Dreyer}.} \bibinfo{year}{2015}\natexlab{}.
\newblock \showarticletitle{Iris: Monoids and Invariants as an Orthogonal Basis
  for Concurrent Reasoning}. In \bibinfo{booktitle}{\emph{Proc. 42nd ACM
  Symposium on Principles of Programming Languages (POPL'15)}}.
  \bibinfo{pages}{637--650}.
\newblock


\bibitem[Kildall(1973)]{kildall}
\bibfield{author}{\bibinfo{person}{Gary~A. Kildall}.}
  \bibinfo{year}{1973}\natexlab{}.
\newblock \showarticletitle{A unified approach to global program optimization}.
  In \bibinfo{booktitle}{\emph{Proceedings of the 1st Annual ACM SIGACT-SIGPLAN
  Symposium on Principles of Programming Languages}} (Boston, Massachusetts)
  \emph{(\bibinfo{series}{POPL '73})}. \bibinfo{publisher}{Association for
  Computing Machinery}, \bibinfo{address}{New York, NY, USA},
  \bibinfo{pages}{194–206}.
\newblock
\showISBNx{9781450373494}
\href{https://doi.org/10.1145/512927.512945}{doi:\nolinkurl{10.1145/512927.512945}}


\bibitem[Lattuada et~al\mbox{.}(2023)]{Verus}
\bibfield{author}{\bibinfo{person}{Andrea Lattuada}, \bibinfo{person}{Travis
  Hance}, \bibinfo{person}{Chanhee Cho}, \bibinfo{person}{Matthias Brun},
  \bibinfo{person}{Isitha Subasinghe}, \bibinfo{person}{Yi Zhou},
  \bibinfo{person}{Jon Howell}, \bibinfo{person}{Bryan Parno}, {and}
  \bibinfo{person}{Chris Hawblitzel}.} \bibinfo{year}{2023}\natexlab{}.
\newblock \showarticletitle{Verus: Verifying Rust Programs using Linear Ghost
  Types}.
\newblock \bibinfo{journal}{\emph{Proc. ACM Program. Lang.}}
  \bibinfo{volume}{7}, \bibinfo{number}{OPSLA1}, Article
  \bibinfo{articleno}{85} (\bibinfo{date}{April} \bibinfo{year}{2023}),
  \bibinfo{numpages}{30}~pages.
\newblock
\href{https://doi.org/10.1145/3586037}{doi:\nolinkurl{10.1145/3586037}}


\bibitem[Lehmann et~al\mbox{.}(2023)]{Flux}
\bibfield{author}{\bibinfo{person}{Nico Lehmann}, \bibinfo{person}{Adam~T.
  Geller}, \bibinfo{person}{Niki Vazou}, {and} \bibinfo{person}{Ranjit Jhala}.}
  \bibinfo{year}{2023}\natexlab{}.
\newblock \showarticletitle{Flux: Liquid Types for Rust}.
\newblock \bibinfo{journal}{\emph{Proc. ACM Program. Lang.}}
  \bibinfo{volume}{7}, \bibinfo{number}{PLDI}, Article \bibinfo{articleno}{169}
  (\bibinfo{date}{June} \bibinfo{year}{2023}), \bibinfo{numpages}{25}~pages.
\newblock
\href{https://doi.org/10.1145/3591283}{doi:\nolinkurl{10.1145/3591283}}


\bibitem[Leroy(2009a)]{leroy09}
\bibfield{author}{\bibinfo{person}{Xavier Leroy}.}
  \bibinfo{year}{2009}\natexlab{a}.
\newblock \showarticletitle{Formal Verification of a Realistic Compiler}.
\newblock \bibinfo{journal}{\emph{Commun. ACM}} \bibinfo{volume}{52},
  \bibinfo{number}{7} (\bibinfo{year}{2009}), \bibinfo{pages}{107--115}.
\newblock
\href{https://doi.org/10.1145/1538788.1538814}{doi:\nolinkurl{10.1145/1538788.1538814}}


\bibitem[Leroy(2009b)]{Leroy-backend}
\bibfield{author}{\bibinfo{person}{Xavier Leroy}.}
  \bibinfo{year}{2009}\natexlab{b}.
\newblock \showarticletitle{A Formally Verified Compiler Back-end}.
\newblock \bibinfo{journal}{\emph{Journal of Automated Reasoning}}
  \bibinfo{volume}{43}, \bibinfo{number}{4} (\bibinfo{year}{2009}),
  \bibinfo{pages}{363--446}.
\newblock
\href{https://doi.org/10.1007/s10817-009-9155-4}{doi:\nolinkurl{10.1007/s10817-009-9155-4}}


\bibitem[Leroy et~al\mbox{.}(2012)]{compcert-mem-v2}
\bibfield{author}{\bibinfo{person}{Xavier Leroy}, \bibinfo{person}{Andrew~W.
  Appel}, \bibinfo{person}{Sandrine Blazy}, {and} \bibinfo{person}{Gordon
  Stewart}.} \bibinfo{year}{2012}\natexlab{}.
\newblock \bibinfo{booktitle}{\emph{{The CompCert Memory Model, Version 2}}}.
\newblock \bibinfo{type}{Research Report} RR-7987.
  \bibinfo{institution}{{INRIA}}. \bibinfo{pages}{26} pages.
\newblock
\urldef\tempurl \url{https://hal.inria.fr/hal-00703441}
\showURL{\tempurl}


\bibitem[Matsakis(2017)]{two-phase-borrow}
\bibfield{author}{\bibinfo{person}{Niko Matsakis}.}
  \bibinfo{year}{2017}\natexlab{}.
\newblock \bibinfo{title}{{RFC 2025}: Enable Nested Method Calls}.
\newblock \bibinfo{howpublished}{The Rust RFC Book}.
\newblock
\urldef\tempurl \url{https://rust-lang.github.io/rfcs/2025-nested-method-calls.html}
\showURL{\tempurl}


\bibitem[Matsakis(2018)]{polonius-alias}
\bibfield{author}{\bibinfo{person}{Niko Matsakis}.}
  \bibinfo{year}{2018}\natexlab{}.
\newblock \bibinfo{title}{An alias-based formulation of the borrow checker}.
\newblock
\urldef\tempurl \url{https://smallcultfollowing.com/babysteps/blog/2018/04/27/an-alias-based-formulation-of-the-borrow-checker/}
\showURL{\tempurl}


\bibitem[Matsakis(2019)]{subset-error}
\bibfield{author}{\bibinfo{person}{Niko Matsakis}.}
  \bibinfo{year}{2019}\natexlab{}.
\newblock \bibinfo{title}{Polonius and region errors}.
\newblock
\urldef\tempurl \url{https://smallcultfollowing.com/babysteps//blog/2019/01/17/polonius-and-region-errors/}
\showURL{\tempurl}


\bibitem[Matsushita et~al\mbox{.}(2021)]{RustHorn}
\bibfield{author}{\bibinfo{person}{Yusuke Matsushita}, \bibinfo{person}{Takeshi
  Tsukada}, {and} \bibinfo{person}{Naoki Kobayashi}.}
  \bibinfo{year}{2021}\natexlab{}.
\newblock \showarticletitle{RustHorn: CHC-based Verification for Rust
  Programs}.
\newblock \bibinfo{journal}{\emph{ACM Trans. Program. Lang. Syst.}}
  \bibinfo{volume}{43}, \bibinfo{number}{4}, Article \bibinfo{articleno}{15}
  (\bibinfo{date}{Oct.} \bibinfo{year}{2021}), \bibinfo{numpages}{54}~pages.
\newblock
\showISSN{0164-0925}
\href{https://doi.org/10.1145/3462205}{doi:\nolinkurl{10.1145/3462205}}


\bibitem[Nowacki et~al\mbox{.}(2026)]{move-borrow-checker}
\bibfield{author}{\bibinfo{person}{Todd Nowacki}, \bibinfo{person}{Sam
  Blackshear}, \bibinfo{person}{John Mitchell}, \bibinfo{person}{Shaz Qadeer},
  {and} \bibinfo{person}{Ilya Sergey}.} \bibinfo{year}{2026}\natexlab{}.
\newblock \showarticletitle{Tracking Borrows with Regular Expressions}.
\newblock \bibinfo{journal}{\emph{Proc. ACM Program. Lang.}}
  \bibinfo{volume}{10}, \bibinfo{number}{OOPSLA2}, Article
  \bibinfo{articleno}{389} (\bibinfo{date}{Oct.} \bibinfo{year}{2026}),
  \bibinfo{numpages}{27}~pages.
\newblock
\href{https://doi.org/10.1145/3839521}{doi:\nolinkurl{10.1145/3839521}}


\bibitem[Ogawa et~al\mbox{.}(2025)]{Thrust}
\bibfield{author}{\bibinfo{person}{Hiromi Ogawa}, \bibinfo{person}{Taro
  Sekiyama}, {and} \bibinfo{person}{Hiroshi Unno}.}
  \bibinfo{year}{2025}\natexlab{}.
\newblock \showarticletitle{Thrust: A Prophecy-Based Refinement Type System for
  Rust}.
\newblock \bibinfo{journal}{\emph{Proc. ACM Program. Lang.}}
  \bibinfo{volume}{9}, \bibinfo{number}{PLDI}, Article \bibinfo{articleno}{230}
  (\bibinfo{date}{June} \bibinfo{year}{2025}), \bibinfo{numpages}{25}~pages.
\newblock
\href{https://doi.org/10.1145/3729333}{doi:\nolinkurl{10.1145/3729333}}


\bibitem[{OpenAI}(2026)]{codex}
\bibfield{author}{\bibinfo{person}{{OpenAI}}.} \bibinfo{year}{2026}\natexlab{}.
\newblock \bibinfo{title}{Codex: A Lightweight Coding Agent That Runs in Your
  Terminal}.
\newblock \bibinfo{howpublished}{\url{https://github.com/openai/codex}}.
\newblock


\bibitem[Pearce(2021)]{light-rust}
\bibfield{author}{\bibinfo{person}{David~J. Pearce}.}
  \bibinfo{year}{2021}\natexlab{}.
\newblock \showarticletitle{A Lightweight Formalism for Reference Lifetimes and
  Borrowing in Rust}.
\newblock \bibinfo{journal}{\emph{ACM Trans. Program. Lang. Syst.}}
  \bibinfo{volume}{43}, \bibinfo{number}{1}, Article \bibinfo{articleno}{3}
  (\bibinfo{date}{April} \bibinfo{year}{2021}), \bibinfo{numpages}{73}~pages.
\newblock
\showISSN{0164-0925}
\href{https://doi.org/10.1145/3443420}{doi:\nolinkurl{10.1145/3443420}}


\bibitem[{pnkfelix}(2018)]{rustc-drop-box}
\bibfield{author}{\bibinfo{person}{{pnkfelix}}.}
  \bibinfo{year}{2018}\natexlab{}.
\newblock \bibinfo{title}{[NLL] Dangly paths for box \#52782}.
\newblock
\urldef\tempurl \url{https://github.com/rust-lang/rust/pull/52782}
\showURL{\tempurl}


\bibitem[Reynaud~Michez(2026)]{sound-borrow-formalize}
\bibfield{author}{\bibinfo{person}{Alban Reynaud~Michez}.}
  \bibinfo{year}{2026}\natexlab{}.
\newblock \showarticletitle{Formal Verification of Borrow-Checking by Local
  Commutation Diagrams}. In \bibinfo{booktitle}{\emph{JFLA 2026 -- 37es
  Journ\'ees Francophones des Langages Applicatifs}},
  \bibfield{editor}{\bibinfo{person}{Marie Kerjean} {and}
  \bibinfo{person}{Yannick Zakowski}} (Eds.). \bibinfo{address}{Oberbronn,
  Alsace, France}.
\newblock
\urldef\tempurl \url{https://hal.science/hal-05428143}
\showURL{\tempurl}
\newblock
\shownote{hal-05428143v2}.


\bibitem[{Rust Compiler Team}(2015)]{mir-rfc}
\bibfield{author}{\bibinfo{person}{{Rust Compiler Team}}.}
  \bibinfo{year}{2015}\natexlab{}.
\newblock \bibinfo{title}{The Rust RFC Book: 1211-mir}.
\newblock
\urldef\tempurl \url{https://rust-lang.github.io/rfcs/1211-mir.html}
\showURL{\tempurl}


\bibitem[{Rust Compiler Team}(2026)]{move-paths}
\bibfield{author}{\bibinfo{person}{{Rust Compiler Team}}.}
  \bibinfo{year}{2026}\natexlab{}.
\newblock \bibinfo{title}{Move paths}.
\newblock
\urldef\tempurl \url{https://rustc-dev-guide.rust-lang.org/borrow-check/moves-and-initialization/move-paths.html}
\showURL{\tempurl}


\bibitem[{Rust contributors}(2019)]{rust-polonius}
\bibfield{author}{\bibinfo{person}{{Rust contributors}}.}
  \bibinfo{year}{2019}\natexlab{}.
\newblock \bibinfo{title}{Defines the Rust borrow checker}.
\newblock
\urldef\tempurl \url{https://github.com/rust-lang/polonius}
\showURL{\tempurl}


\bibitem[{Rust contributors}(2021)]{forward-polonius}
\bibfield{author}{\bibinfo{person}{{Rust contributors}}.}
  \bibinfo{year}{2021}\natexlab{}.
\newblock \bibinfo{title}{polonius.next: experimental datalog rules for a next
  gen polonius}.
\newblock
\urldef\tempurl \url{https://github.com/nikomatsakis/polonius.next/tree/reintroduce-leases}
\showURL{\tempurl}


\bibitem[{Rust contributors}(2026a)]{a-mir-formality}
\bibfield{author}{\bibinfo{person}{{Rust contributors}}.}
  \bibinfo{year}{2026}\natexlab{a}.
\newblock \bibinfo{title}{a-mir-formality: a model of MIR and the Rust
  type/trait system}.
\newblock
\urldef\tempurl \url{https://github.com/rust-lang/a-mir-formality}
\showURL{\tempurl}


\bibitem[{Rust contributors}(2026b)]{rustc-polonius}
\bibfield{author}{\bibinfo{person}{{Rust contributors}}.}
  \bibinfo{year}{2026}\natexlab{b}.
\newblock \bibinfo{title}{Implementation of Polonius in Rust compiler}.
\newblock
\urldef\tempurl \url{https://github.com/rust-lang/rust/tree/main/compiler/rustc_borrowck/src/polonius}
\showURL{\tempurl}


\bibitem[{Rust contributors}(2026c)]{rustc-tests}
\bibfield{author}{\bibinfo{person}{{Rust contributors}}.}
  \bibinfo{year}{2026}\natexlab{c}.
\newblock \bibinfo{title}{The Rust compiler test suites}.
\newblock
\urldef\tempurl \url{https://github.com/rust-lang/rust/tree/main/tests/ui}
\showURL{\tempurl}


\bibitem[{Rémy Rakic and Niko Matsakis on behalf of The Polonius Working
  Group}(2023)]{Polonius-update}
\bibfield{author}{\bibinfo{person}{{Rémy Rakic and Niko Matsakis on behalf of
  The Polonius Working Group}}.} \bibinfo{year}{2023}\natexlab{}.
\newblock \bibinfo{title}{Polonius Update}.
\newblock
\urldef\tempurl \url{https://blog.rust-lang.org/inside-rust/2023/10/06/polonius-update.html}
\showURL{\tempurl}


\bibitem[{soulwa}(2020)]{rust-auto-reborrow}
\bibfield{author}{\bibinfo{person}{{soulwa}}.} \bibinfo{year}{2020}\natexlab{}.
\newblock \bibinfo{title}{Better documentation of reborrowing}.
\newblock \bibinfo{howpublished}{GitHub issue \#788, rust-lang/reference}.
\newblock
\urldef\tempurl \url{https://github.com/rust-lang/reference/issues/788}
\showURL{\tempurl}


\bibitem[Stjerna(2020)]{Polonius-thesis}
\bibfield{author}{\bibinfo{person}{Amanda Stjerna}.}
  \bibinfo{year}{2020}\natexlab{}.
\newblock \emph{\bibinfo{title}{Modelling Rust’s Reference Ownership Analysis
  Declaratively in Datalog}}.
\newblock \bibinfo{thesistype}{Master's\ thesis}. \bibinfo{school}{Uppsala
  University, Department of Information Technology}.
\newblock


\bibitem[Team(2017)]{rust-nll}
\bibfield{author}{\bibinfo{person}{The~Rust Team}.}
  \bibinfo{year}{2017}\natexlab{}.
\newblock \bibinfo{title}{Non-Lexical Lifetimes}.
\newblock
\urldef\tempurl \url{https://rust-lang.github.io/rfcs/2094-nll.html}
\showURL{\tempurl}


\bibitem[{The Chromium Projects}(0000)]{chromium-memory-safety}
\bibfield{author}{\bibinfo{person}{{The Chromium Projects}}.}
  \bibinfo{year}{[n.\,d.]}\natexlab{}.
\newblock \bibinfo{title}{Memory Safety}.
\newblock \bibinfo{howpublished}{Chromium Security}.
\newblock
\urldef\tempurl \url{https://www.chromium.org/Home/chromium-security/memory-safety/}
\showURL{\tempurl}


\bibitem[{The Rust Team}(2025)]{rust-subtyping}
\bibfield{author}{\bibinfo{person}{{The Rust Team}}.}
  \bibinfo{year}{2025}\natexlab{}.
\newblock \bibinfo{title}{Subtyping and Variance}.
\newblock
\urldef\tempurl \url{https://doc.rust-lang.org/nomicon/subtyping.html}
\showURL{\tempurl}


\bibitem[{The Rust Team}(2026)]{rust}
\bibfield{author}{\bibinfo{person}{{The Rust Team}}.}
  \bibinfo{year}{2026}\natexlab{}.
\newblock \bibinfo{title}{The Rust Programming Language}.
\newblock
\urldef\tempurl \url{https://rust-lang.org/}
\showURL{\tempurl}


\bibitem[Thomas(2019)]{microsoft-memory-safety}
\bibfield{author}{\bibinfo{person}{Gavin Thomas}.}
  \bibinfo{year}{2019}\natexlab{}.
\newblock \bibinfo{title}{A Proactive Approach to More Secure Code}.
\newblock \bibinfo{howpublished}{Microsoft Security Response Center Blog}.
\newblock
\urldef\tempurl \url{https://www.microsoft.com/en-us/msrc/blog/2019/07/a-proactive-approach-to-more-secure-code}
\showURL{\tempurl}


\bibitem[Villani et~al\mbox{.}(2025)]{tree-borrow}
\bibfield{author}{\bibinfo{person}{Neven Villani}, \bibinfo{person}{Johannes
  Hostert}, \bibinfo{person}{Derek Dreyer}, {and} \bibinfo{person}{Ralf Jung}.}
  \bibinfo{year}{2025}\natexlab{}.
\newblock \showarticletitle{Tree Borrows}.
\newblock \bibinfo{journal}{\emph{Proc. ACM Program. Lang.}}
  \bibinfo{volume}{9}, \bibinfo{number}{PLDI}, Article \bibinfo{articleno}{188}
  (\bibinfo{date}{June} \bibinfo{year}{2025}), \bibinfo{numpages}{24}~pages.
\newblock
\href{https://doi.org/10.1145/3735592}{doi:\nolinkurl{10.1145/3735592}}


\bibitem[Weiss et~al\mbox{.}(2019)]{weiss2019oxide}
\bibfield{author}{\bibinfo{person}{Aaron Weiss}, \bibinfo{person}{Olek
  Gierczak}, \bibinfo{person}{Daniel Patterson}, {and} \bibinfo{person}{Amal
  Ahmed}.} \bibinfo{year}{2019}\natexlab{}.
\newblock \showarticletitle{Oxide: The essence of rust}.
\newblock \bibinfo{journal}{\emph{arXiv preprint arXiv:1903.00982}}
  (\bibinfo{year}{2019}).
\newblock


\end{thebibliography}
\end{document}